\documentclass[format=acmsmall, review=true]{acmart}
\usepackage{acm-ec-25}
\usepackage{booktabs} 
\usepackage[ruled]{algorithm2e} 
\usepackage{longtable}

\usepackage{epsfig}
\usepackage{ifsym}
\usepackage{ifthen}
\usepackage{verbatim}
\usepackage{color}
\usepackage{amsthm}
\usepackage{xcolor}
\usepackage{graphicx}
\usepackage{bbm}
\usepackage{hyperref}
\hypersetup{colorlinks=true, citecolor=blue, linkcolor=red, urlcolor=pink, pdfstartview={FitH}}
\usepackage{xspace}
\usepackage{natbib}
\usepackage{balance}

\usepackage{seqsplit}

\usepackage[capitalise]{cleveref}

\usepackage{array}   
\usepackage{multirow}
\newcolumntype{C}[1]{>{\centering\arraybackslash}m{#1}}

\newtheorem{prop}{Proposition}

\usepackage{array}
\usepackage{subcaption}
\SetAlFnt{\small}
\SetAlCapFnt{\small}
\SetAlCapNameFnt{\small}
\SetAlCapHSkip{0pt}
\IncMargin{-\parindent}
\usepackage{amsmath}

\makeatletter
\long\def\tlist@if@empty@nTF #1{%
\expandafter\ifx\expandafter\\\detokenize{#1}\\%
\expandafter\@firstoftwo
\else
\expandafter\@secondoftwo
\fi
}

\newcommand{\capcap}{\textsl}
\newcommand{\capcapcap}{}

\defcitealias{dworczak2024optimal}{Dworczak \textsuperscript{\textcircled{r}}\ Reuter \textsuperscript{\textcircled{r}}\ Kominers \textsuperscript{\textcircled{r}}\ Lee}

\usepackage{longtable}
\usepackage[tableposition=below]{caption}
\usepackage{amsmath,amsthm}
\usepackage{thmtools}
\declaretheoremstyle[
spaceabove=6pt, spacebelow=6pt,
headfont=\normalfont\bfseries,
notefont=\mdseries, notebraces={(}{)},
bodyfont=\normalfont,
postheadspace=0.6em,
headpunct=:
]{hypstyle}

\theoremstyle{definition}

\crefname{hyp}{hypothesis}{hypotheses}
\Crefname{hyp}{Hypothesis}{Hypotheses}

\newif\ifcomments
\commentstrue
\ifcomments
    \providecommand{\narun}[1]{{\protect\color{blue}{[Narun: #1]}}}
    \providecommand{\tl}[1]{{\protect\color{magenta}{[Taylor: #1]}}} 
    \providecommand{\klb}[1]{{\protect\color{red}{[Kevin: #1]}}} 
    \providecommand{\sdk}[1]{{\protect\color{orange}{[sk: #1]}}} 
\else
    \providecommand{\narun}[1]{}
    \providecommand{\tl}[1]{}
    \providecommand{\klb}[1]{} 
    \providecommand{\sdk}[1]{} 
\fi

\newcommand{\nc}{\newcommand}
\nc\zmin{z_{\min}}
\nc\zmax{z_{\max}}

\nc\soc[1][]{\Sigma\tlist@if@empty@nTF{#1}{}{(#1)}}
\nc\nx{x_{-i}}
\nc\xnx{x;\nx}
\nc\coa{\alpha}
\nc\cob{\beta}
\nc\coc{\gamma}
\nc\Nfn[1][]{N\tlist@if@empty@nTF{#1}{}{(#1)}}
\nc\Nfnx[1][]{N_x\tlist@if@empty@nTF{#1}{}{(#1)}}
\nc\Nfnxj[1][]{N_{xx_j}\tlist@if@empty@nTF{#1}{}{(#1)}}
\nc\Nfnxx[1][]{N_{xx}\tlist@if@empty@nTF{#1}{}{(#1)}}
\nc\Dfn[1][]{D\tlist@if@empty@nTF{#1}{}{(#1)}}
\nc\Afn[1][]{A\tlist@if@empty@nTF{#1}{}{(#1)}}
\nc\dfn[1][]{d\tlist@if@empty@nTF{#1}{}{(#1)}}
\nc\afn[1][]{a\tlist@if@empty@nTF{#1}{}{(#1)}}
\nc\optx[1][]{x^*\tlist@if@empty@nTF{#1}{}{(#1)}}

\title[NFTs as a Data-Rich Test Bed: Conspicuous Consumption and its Determinants]{NFTs as a Data-Rich Test Bed: \\Conspicuous Consumption and its Determinants}

\author{Taylor Lundy}
\affiliation{%
  \institution{University of British Columbia}
  \city{Vancouver}
  \state{BC}
  \country{Canada}}
\email{tlundy@cs.ubc.ca}
\author{Narun Raman}
\affiliation{%
  \institution{University of British Columbia}
  \city{Vancouver}
  \state{BC}
  \country{Canada}}
\email{narunram@cs.ubc.ca}
\author{Scott Duke Kominers}
\affiliation{%
  \institution{Harvard University \& a16z crypto}
  \city{Cambridge}
  \state{MA}
  \country{USA}}
\email{kominers@fas.harvard.edu}

\author{Kevin Leyton-Brown}
\affiliation{%
  \institution{University of British Columbia}
  \city{Vancouver}
  \state{BC}
  \country{Canada}}
\email{kevinlb@cs.ubc.ca}

\begin{abstract}
Conspicuous consumption occurs when a consumer derives value from a good based on its social meaning as a signal of wealth, taste, and/or community affiliation.
Common conspicuous goods include designer footwear, country club memberships, and artwork; conspicuous goods also exist in the digital sphere, with non-fungible tokens (NFTs) as a prominent example.
The NFT market merits deeper study for two key reasons: first, it is poorly understood relative to its economic scale; and second, it is unusually amenable to analysis because NFT transactions are publicly available on the blockchain, making them useful as a test bed for conspicuous consumption dynamics.
This paper introduces a model that incorporates two previously identified elements of conspicuous consumption: the \emph{bandwagon effect} (goods increase in value as they become more popular) and the \emph{snob effect} (goods increase in value as they become rarer). Our model resolves the apparent tension between these two effects, exhibiting net complementarity between others' and one's own conspicuous consumption. We also introduce a novel dataset combining NFT transactions with embeddings of the corresponding NFT images computed using an off-the-shelf vision transformer architecture. We use our dataset to validate the model, showing that the bandwagon effect raises an NFT collection's value as more consumers join, while the snob effect drives consumers to seek rarer NFTs within a given collection.
\end{abstract}

\begin{document}

\maketitle

\section{Introduction}

NFTs are crypto\-graph\-ically-secured records of ownership, allowing digital goods like images and other media files to be certifiably owned, and thus exchanged. Amid the emerging popularity of NFTs there is confusion about their purpose among both consumers and researchers. Headlines in the early days of NFTs sometimes referred to them as scams or ``pyramid schemes'' \cite{axie2022}, making it harder for consumers to see their legitimate value. Additionally, previous studies often focused on NFTs as investment assets \citep{wang2023dissecting,ko2022economic}, which fails to explain behaviors such as NFTs being given away for free to prominent members of the crypto industry to create hype (see, e.g., \cite{nakamigos2023}) or the formation of communities with Discord channels, $\mathbb{X}$ accounts, and forums, where holders of the NFTs interact with each other (see, e.g., \cite{kaczynski2023everything}). 
Instead, these behaviors imply that owners derive significant value from NFTs by using them to signal both status (flaunting something rare) and establish community affiliation (using aesthetic choices and social connections to adjudicate membership in a group). 

Indeed, one of the most popular categories of NFTs, \emph{profile picture} (\emph{PFP}) NFTs,\footnote{``PFP'' technically stands for ``\emph{p}icture \emph{f}or \emph{p}roof,'' but in colloquial parlance has also been taken to mean ``\emph{p}ro\emph{f}ile \emph{p}icture.''} are associated with images that are intended to serve as online personae on social networks like $\mathbb{X}$ (fka.~Twitter), Facebook, and Farcaster; other NFT categories include those conveying ownership of ``skins'' or items for gaming characters, other digital wearables, or fine art. 
In other words, NFTs are ``conspicuous goods''; their value derives not only from their function but also from the social meaning they convey.

A variety of consumer behaviors have been identified in the literature on conspicuous goods. In 1899, Thorstein Veblen first argued that consumer demand for certain goods and services arises from a desire to establish social affiliations and emulate higher social classes and economic groups \citep{veblen1955}. Since then, distinct status-seeking consumption patterns have been found in a variety of contexts \citep{leibenstein1950bandwagon, chaudhuri2006diamonds, shukla2008conspicuous, gierl2010scarce, bekir2013luxury, chen2016research}.  Notably, \citet{leibenstein1950bandwagon} identified two effects that influence the utility derived from conspicuous goods: the ``bandwagon effect,'' whereby demand for a luxury good increases as more consumers consume it, and the ``snob effect,'' whereby demand increases as a good becomes rarer. 
Subsequent work has identified social dynamics that create and reinforce these effects. \citet{vigneron1999review} found that conforming with aspirational groups and a desire to be fashionable are primary drivers for the bandwagon effect. \citet{han2010signaling} showed that social structures can significantly influence consumer preferences, with higher-income consumers preferring subtler status signals recognizable only within their social circles; \citet{carbajal2015inconspicuous} showed furthermore that this effect might be most prominent in highly socially connected ``old money'' individuals. Research has also shown that the snob effect can be more pronounced between direct acquaintances \citep{kuwashima2016structural}.  Finally, these social dynamics are not limited to physical conspicuous goods, as researchers increasingly account for them when evaluating and pricing digital goods \citep{geng2019optimal,goetz2022peer,lundy2024pay}. 

NFTs give us access to a much richer dataset than has previously existed for conspicuous goods: the NFTs we examine are recorded on a publicly readable blockchain ledger, making both the associated digital goods themselves (often images) and their transaction data (who bought; who sold; and the transaction price) globally accessible. Researchers have already begun to leverage the data NFTs provide to learn not only about the NFT market itself \citep{nadini2021mapping} but also the forces underlying it. For example, \citet{digitalveblen2022} have shown that during their primary sales, NFTs act as Veblen goods, a type of conspicuous good, and are influenced by the bandwagon effect. Notably, in the primary market where new NFTs are initially ``minted,'' it is relatively straightforward to identify which collections are popular based on their sell-out rate: collections (endogenously) determined to be trendy sell out entirely, while most others see minimal sales.\footnote{\citet{dworczak2024optimal} introduced a framework of ``optimal membership design'' that covers the design of networks with cross-member externalities, including NFT communities; the externalities studied there can include conspicuous consumption dynamics such as bandwagon and snob effects.}

This paper begins by extending \citet{hopkins2024cardinal} to build a model (\cref{sec:model}) that captures both the bandwagon effect and the snob effect. Our model allows these two effects to coexist by being sensitive to different levels of consumption.  Specifically, the bandwagon effect dominates as new consumers purchase goods of a given type (an NFT collection; a designer brand; etc), causing its overall value to increase with popularity.  Within such types of goods, however, the snob effect becomes the driving force, pushing consumers to acquire increasingly rare items to stand out from the crowd. 
To validate and explore the implications of our model, we constructed a novel dataset of image-based NFTs (\cref{sec:dataset}), including 48,595,074 NFTs organized into 10,963 collections, and the 3,755,256 unique ``wallets'' that hold them. Our dataset also goes beyond transaction data, including a subset of images for each NFT collection; this allows us to quantify the visual similarity both between individual NFTs and between NFT collections, using a pre-trained vision transformer network. 
Analyzing this data, we demonstrate (\cref{sec:bandwagon}) that collection floor prices reflect the bandwagon effect, and identify two key features that drive this effect within the NFT market: community affiliation and wealth. We also show that the bandwagon effect depends not only on the power of influencers but on a network effect among the less influential members of the community. We then show (\cref{sec:snob}) that variation in prices at the NFT level within-collection reflect the snob effect, and compare the relative power rarity and visual distinctiveness have in affecting sale prices. Finally, in \cref{sec:conc}, we discuss the broader value of our NFT dataset and some potential future work.

\section{Model}\label{sec:model}

A formal description of our model appears in \cref{app:formal_model}; here we outline some of its qualitative implications that are relevant for our empirical analysis.

Our model characterizes the consumption decisions of a continuum of agents for a conspicuous good under utility functions that depend both directly on the level at which an individual consumes that good and on a social interaction term reflecting the good's overall pattern of consumption in the population.
The social interaction term captures both a bandwagon effect, implemented through a positive network externality of consuming the conspicuous good, and a snob effect, implemented through a preference for consuming the conspicuous good at a higher level than others.

\Cref{prop:1} in \cref{app:formal_model} resolves the apparent tension between the bandwagon and snob effects by showing that even in the presence of the snob effect, there is overall complementarity between others' and one's own consumption of the conspicuous good. Moreover, assuming concavity in the impact of network externalities, the bandwagon and snob effects in a sense operate at different levels of resolution: the impact of one agent $j$'s decision to increase conspicuous consumption on the choices of agents $i\neq j$ is more driven by the bandwagon effect at low levels of $j$'s consumption, and by the snob effect at higher levels of $j$'s consumption. This is consistent with research  suggesting that these two effects can coexist by varying in strength as a function of social proximity; for example,  \citet{kuwashima2016structural} has shown that the snob effect is stronger between those who are direct acquaintances.

A final qualitative conclusion follows from inspection of the proof of \cref{prop:1}: the effect of $j$'s network externality on $i$'s conspicuous consumption is magnified if $j$ is particularly prominent or otherwise valuable to the network. This is consistent with the fact that visible consumption by celebrities or other high-status individuals can particularly increase demand for  conspicuous goods~\cite{belk1988possessions, amaldoss2008research}.

\section{Dataset} \label{sec:dataset}

Our dataset contains information on both individual NFTs and NFT collections. Both the collection and NFT data were collected using a combination of the OpenSea\footnote{https://opensea.io/} and Alchemy NFT\footnote{https://www.alchemy.com/nft-api} APIs. 

\subsubsection*{Metadata}

Our NFT data spans 10,963 image-based NFT collections, i.e., series of NFTs associated with images that are organized into collections via unifying smart contracts. (These collections are reflected in ``collection pages'' on OpenSea.) These collections were mapped using the Opensea API, which automatically filters out ``spam'' collections (such as re-uploads of an existing NFT collection's image assets). In addition, we filter out collections with very large token supplies, such as the collection created by Rarible, which acts as an art exchange and contains over a billion tokens.  Our data contains, for each collection, a list of OpenSea summary transactional data---e.g., current ``floor price'' (the minimum price at which any NFT in the collection is currently listed for direct sale) as of January 2024, total sales volume, and average historical sale price---as well as some additional non-sale related metadata data for each collection---category (e.g., profile picture, art, gaming, and so forth) and creation date (as recorded by OpenSea). ``Profile picture'' (``PFP'') NFTs comprise the largest category in our dataset, representing 4,221 of the collections, the next largest being ``uncategorized,'' with 3,192, and ``art,'' with 2,365. The remaining 1,185 collections are split amongst smaller categories such as ``gaming,'' ``collectibles,'' and ``photography.''

At the NFT level, we store the current owner address of each NFT in each collection as an (owner wallet, token ID) pair, collected on January 26, 2024 (e.g., the owner of Bored Ape \#9976 is recorded in our dataset as Ethereum network address \seqsplit{0x9c7007B750B509dA0c72338de2C2531eD559F4aF}). This gives us 48,595,074 (wallet, token~ID) pairs, associated with a total of 3,755,256 unique wallets.

\subsubsection*{Image Embeddings}

Much of the analysis in this paper relies on computing image similarity and training models; both of which require meaningful image embeddings. We computed embeddings using a pre-trained vision transformer network, DINOv2 \citep{oquab2023dinov2}, which takes in an image and produces a $384$-dimensional real-number embedding. These embeddings have been shown to give state-of-the-art performance at both image retrieval and image clustering tasks with no fine-tuning necessary \citep{oquab2023dinov2}; a similar architecture has also been used in NFT price prediction \citep{costa2023show}.

To retrieve the images, we first pulled and saved the image URLs for all NFTs we wished to retrieve using the OpenSea API. Then, we downloaded the images directly from the source URLs and converted them into $300\times 300$--pixel PNG files.

We also needed an aggregate notion of an embedding for our collection-level data. In each collection, we randomly sampled 50 NFTs and computed the arithmetic mean (or centriod) of their embeddings.\footnote{For a discussion of the stability of these centroids with varying sample sizes, see \cref{app:embed}.} We computed centroids for all 10,963 NFT collections in the dataset.

We would have liked to conduct image analysis at a more fine-grained level. However, it was infeasible to generate image embeddings for every NFT in our dataset, as this would have required retrieving and storing over 50 million images. Thus, we instead created a more fine-grained dataset over a smaller number of collections by randomly sampling 1,265 collections from the main sample.\footnote{We rejected collections from sampling if they did not have sufficient sales data for the downstream analysis we conduct.} For each collection in the subsample, we sampled a total of 600 images per collection, and constructed embeddings as described above, along with centroids based on those 600-image groups. Additionally, we added the average sale price of each individual NFT to this dataset, to aid in analysis we conduct in \cref{sec:snob}.

\subsubsection*{Rarity Ranks}

\begin{figure}[t]
    \centering
    \includegraphics[width=0.4\textwidth]{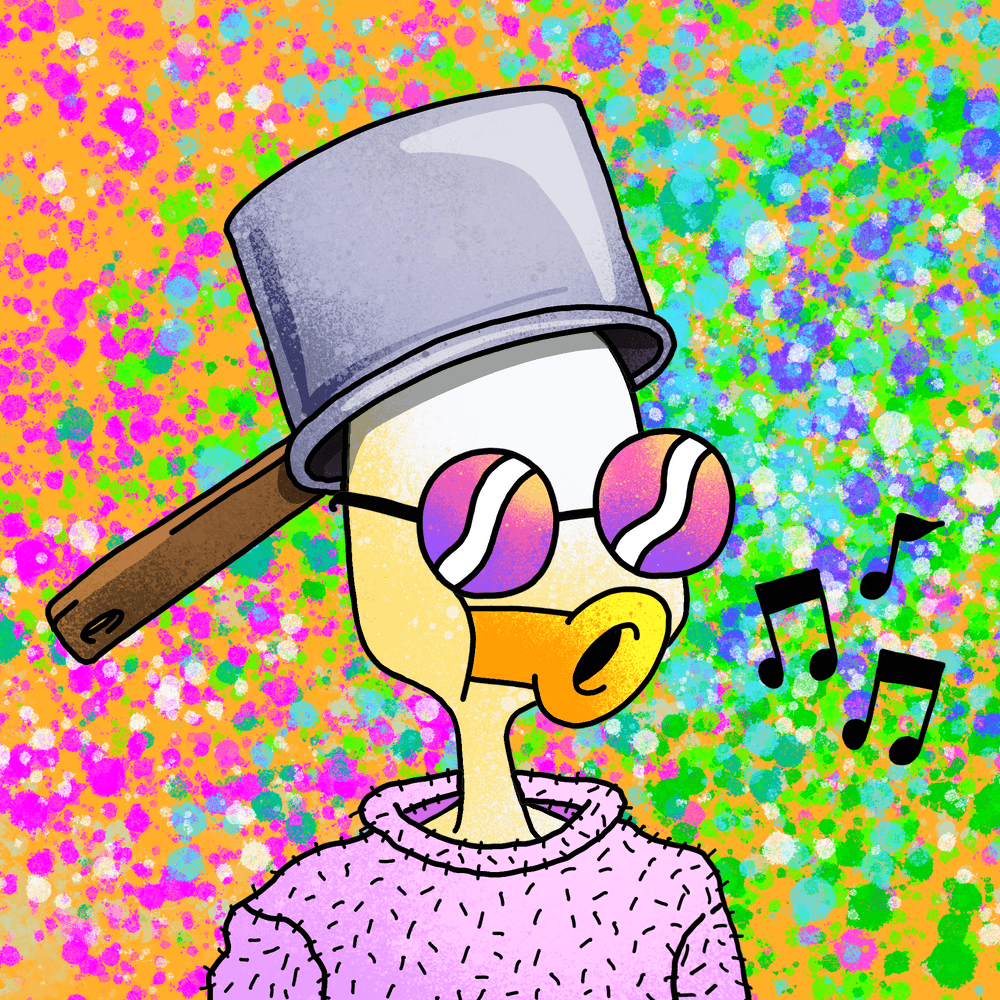} 
    \caption{\capcap{An example of PFP NFT traits.} SupDuck \#6484, image pictured, has the following traits and corresponding rarities: \emph{Background} -- Paint Splatter ($6\%$), \emph{Hat} -- Pot Head ($4\%$), \emph{Clothes} -- Itchy Ass Sweater ($6\%$), \emph{Eyes} -- Hippy ($10\%$), \emph{Mouth} -- Whistle ($2\%$), \emph{Skin} -- Buttermilk ($12\%$). OpenSea (via OpenRarity) ranks SupDuck \#6484 at 5,625 out of 10,001 total SupDucks. }
    \label{fig:supduck-traits}
\end{figure}

Many NFT images are programmatically generated by combining randomly selected visual ``traits.'' Each trait is often associated with a rarity, usually expressed as the proportion of other NFTs in the same collection sharing the trait. For an example, see \cref{fig:supduck-traits}.

While assessing the relative rarity of a given NFT seems like a straightforward task, it is made more complicated by the fact that individual tokens---even within a single collection---may have different numbers of traits, and moreover, it is not always clear how to interpret tokens displaying mixtures of both rare and more common traits. Thus, the NFT community has come up with several different ways to measure the aggregate rarity of an NFT. One commonly used rarity metric is a dense ranking computed using OpenRarity\footnote{https://www.openrarity.dev/}; this is the ranking used by OpenSea, and it ranks NFTs first by the number of traits that only appear once in the collection (i.e., ``one-of-ones'') and then by the information content of the traits. For each of the NFTs in our smaller dataset, we obtained the associated OpenRarity ranks from OpenSea if they were available; of the 1,265 collections, $959$ had valid OpenRarity ranks. 

\subsubsection*{Snob Effect Case Study}
\label{subsubsec:case_study_data}
In addition to the 1,265 collections for which we subsampled images, we also gathered more detailed data for 9 NFT collections with high total sales volumes. We chose these nine collections from the top 30 by sales volume, selecting only those that showed a significant correlation with rarity (see \cref{varsec:case_study} for details). For each of these $9$ collections, we computed an image embedding for every image in the collection. We also gathered the entire transaction history for each token in the collection (i.e., a record of every sale, along with the price, buyer, seller, and timestamp).

\section{The Bandwagon Effect}\label{sec:bandwagon}

In this section, we validate whether NFT values are in part influenced by the bandwagon effect.
Recent work by \citet{digitalveblen2022} showed evidence of a bandwagon effect in the primary-sale NFT market, where popularity is easier to assess because primary sales often follow a bimodal distribution---collections either sell out or yield a relatively small number of sales. The secondary market for NFTs presents a different landscape. Here, nearly every NFT being exchanged has already found an owner, making the task of discerning which collections remain socially desirable more nuanced. Furthermore, what is considered popular or trendy can vary across individuals as a function of which other groups they are affiliated with. We model these social dynamics using an ownership graph which connects each wallet to all of the NFT collections that they hold. The graph consists of two node sets: wallets and NFT collections. An edge connects a wallet node to an NFT collection node if the wallet holds at least one NFT from that collection. We give a more in-depth description of this graph in \Cref{app:bandwagon_exp_setup}.

If the bandwagon effect influences a collection's value, then the ownership graph should have predictive power regarding the value of a collection. Attempting to predict this value using a classical regression model would depend heavily on the features chosen. Thus, we instead use a Graph Neural Network (GNN), training on the ownership graph and asking it to predict the floor price of each of the collections in our dataset. We choose floor price rather than a historical price average because the floor price represents a snapshot of collection value, while our graph represents a snapshot of ownership data. We also explore how the visual characteristics of an NFT might impact its value. Many collections emulate others in order to  take advantage of a trendy aesthetic. To quantify the impact of visual characteristics, we give the GNN access to a collection's centroid as a node feature and measure the extent to which this improves performance. We test whether the bandwagon effect is the predominant force at low levels of consumption by increasing the number of owners for a given collection and measuring whether the predicted floor price increases. Finally, we probe our trained model to get a better understanding of which features it uses to predict value and how they relate to the bandwagon effect. 

\subsection{Measuring Predictive Power}

In order to obtain a sensible prediction target, we first transform floor prices into percentiles, sorting each floor price into one of $100$ buckets based on which percentile of the distribution (over all floor prices) it lands in. This also gives us a sensible baseline: always predicting the median value of all collections.\footnote{We use median rather than the ``$50$'' label because there are point masses in the data.} We defer the remainder of the experimental setup to \cref{app:bandwagon_exp_setup}.

\begin{figure}
    \centering
    \includegraphics[width=0.5\textwidth]{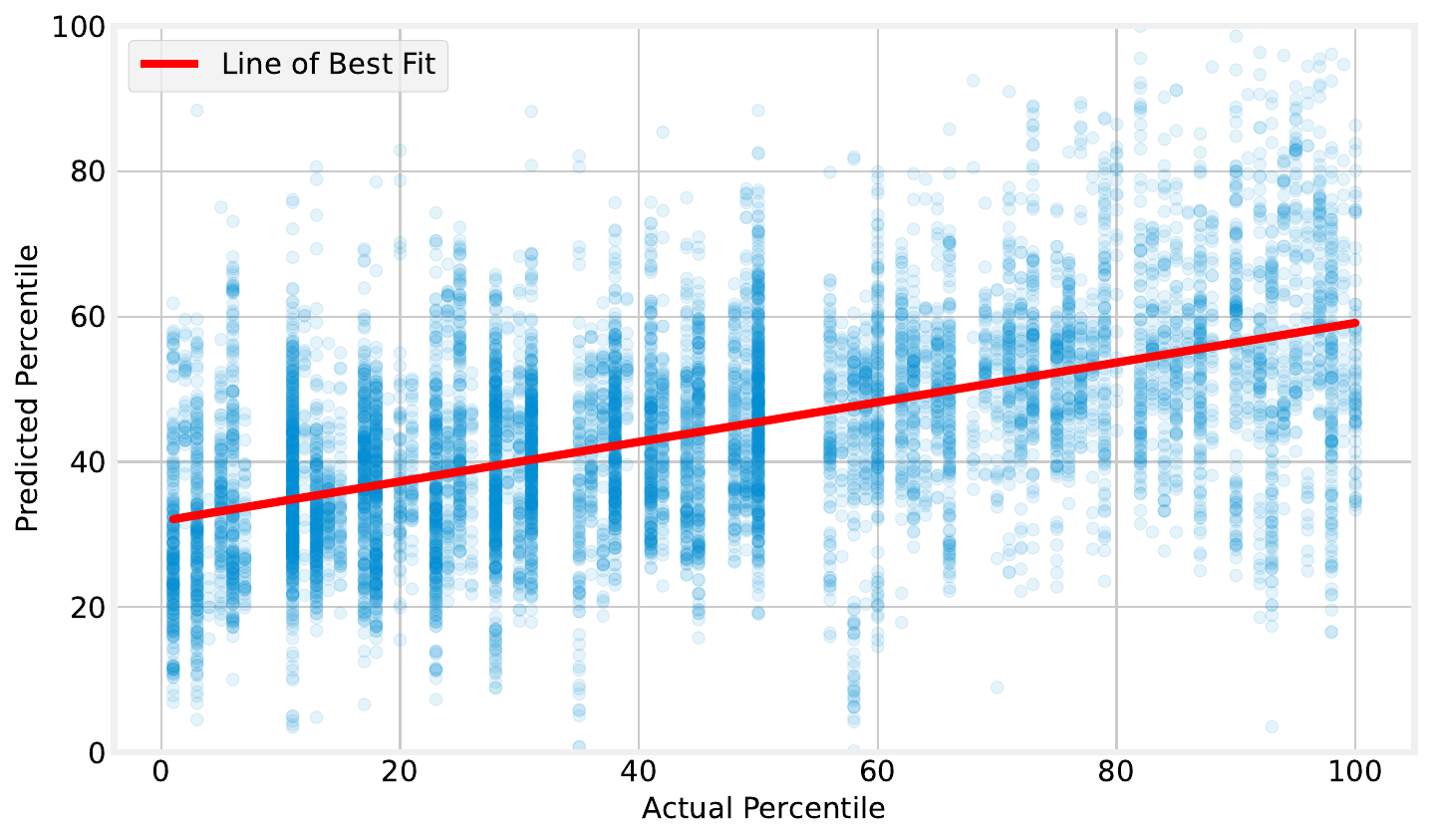}
    \caption{\capcap{Performance of the GNN model with centroids.} This graph illustrates the comparison between the true percentile values of NFT floor prices and those predicted by the model. Each point represents an NFT collection, plotted according to its true percentile in floor price ($x$-axis) against the predicted percentile floor price ($y$-axis).}
    \Description[Performance of the GNN model with centroids.]{This graph illustrates the comparison between the true percentile values of NFT floor prices and those predicted by the model. Each point represents an NFT collection, plotted according to its true percentile in floor price ($x$-axis) against the predicted percentile floor price ($y$-axis).}
    \label{fig:best_model}
\end{figure}

We compare two versions of our graph neural network (GNN) models against the median baseline. We find that the GNN model without centroids outperforms the baseline root mean squared error (RMSE) of $2779$ by $15\%$, achieving a RMSE of $2356$, and the GNN model with centroids outperforms the baseline by $23\%$, with a RMSE of $2133$. Additionally, the predictions of our most effective model, as illustrated in \cref{fig:best_model}, demonstrate a moderate Pearson correlation with the true values: $0.532$ (with $p<0.05$).

\subsection{Evidence of the Bandwagon Effect}
The results so far demonstrate that the ownership graph has predictive power, but it is unclear what aspects of the graph drive these predictions. To address this, we investigate the impact on the model's predictions of adding or removing edges from the graph. If the model has  learned to predict a bandwagon effect, then we should see that adding edges representing ownership of an NFT collection by wallets that are in some sense important or influential should increase the predicted value of that collection;  conversely, removing links to such wallets should lead to a  decrease in predicted value. 

We define the \emph{importance} of a wallet within the graph as the product of two graph properties: wealth and affinity. We define the \emph{wealth} of a wallet node as the sum of the floor prices of all of its connected collections.\footnote{This calculation of wealth does not take into account how many NFTs of each collection a wallet owns. This is because our graphs were constructed unweighted and therefore could not have used a more nuanced notion of wealth to form its predictions.}  We also introduce a notion of the \emph{affinity} of a wallet node, aiming to represent how well its holdings align with the broader NFT-owner community. We begin by identifying the overlap for each collection, i.e., the number of shared wallets for every collection pair. The affinity of a wallet node is then the cumulative overlaps of all its connected collections.

We then modified the ownership graph by first sampling random collections and then, for each collection, sampling non-neighbor wallet nodes to which we added edges, or neighbor wallet nodes from which we deleted edges.\footnote{Due to limitations on the number of nodes we can represent in a GPU, each sample iteration is restricted to specific collection node sets.} We repeated this entire procedure, each time varying the number of edges ($25$, $50$, $100$, or $200$) to add or delete. In the end, we obtained 50,000 samples for each of the number of edges. We also varied the weights in the sampling procedure, sampling by importance, affinity, or wealth, or uniformly over wallet nodes. 

\begin{figure}
    \centering
    \begin{subfigure}{0.49\textwidth}
        \includegraphics[width=\textwidth]{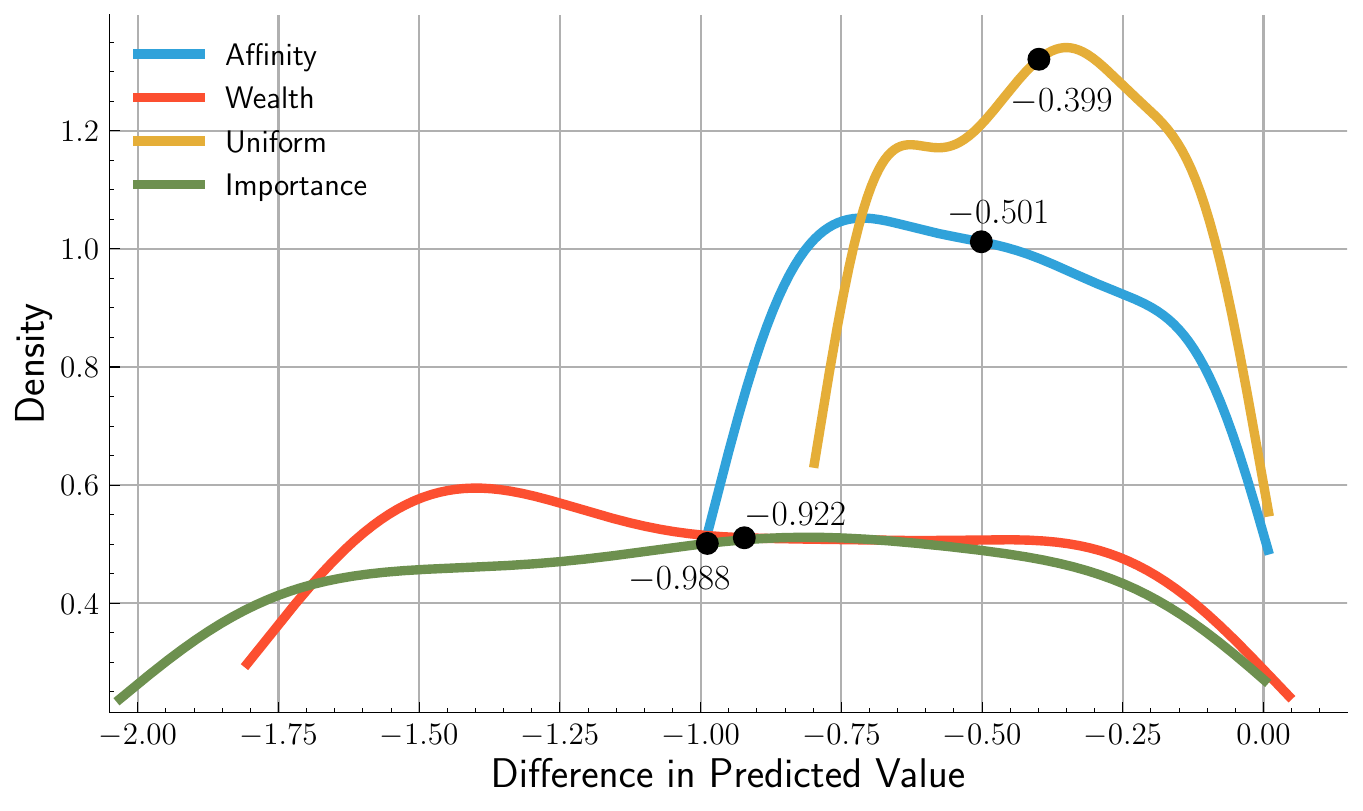}
        \subcaption{\capcapcap{Deleting $100$ edges.}}
        \label{subfig:deleting_edges}
    \end{subfigure}\hfill
    \begin{subfigure}{0.49\textwidth}
        \includegraphics[width=\textwidth]{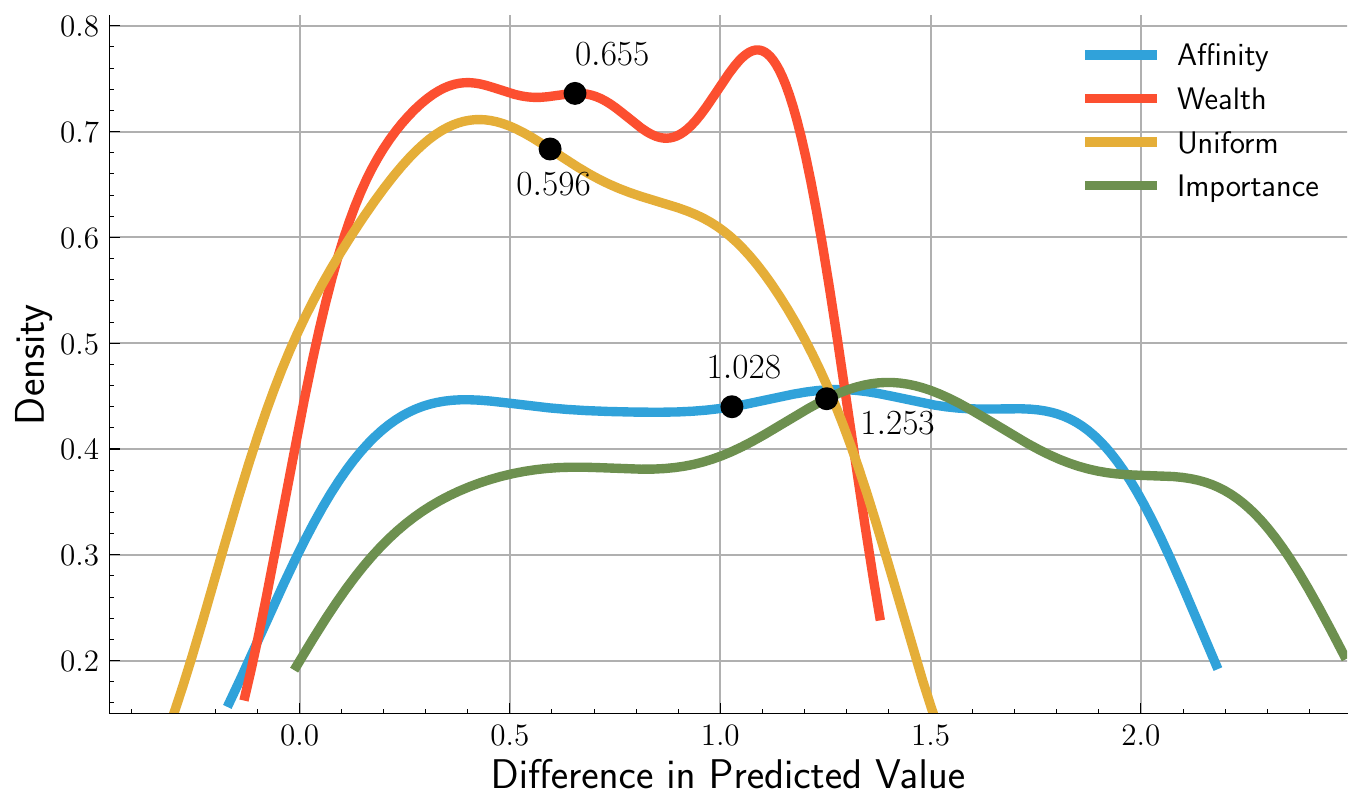}
        \subcaption{\capcapcap{Adding $100$ edges.}}
        \label{subfig:adding_edges}
    \end{subfigure}
    
    \caption{\capcap{Distribution of predicted percentile collection floor price differences on modified graphs, smoothed by kernel density estimation (KDE).} This figure presents KDE plots showing the distribution of differences in predicted values between modified and unmodified graphs. Each line within the plots corresponds to a distinct edge sampling strategy: sampling by affinity (blue), sampling by wealth (red), sampling by importance (green), and uniform sampling (yellow). Means are plotted in black.}
    \Description[Distribution of predicted percentile collection floor price differences on modified graphs, smoothed by kernel density estimation (KDE).]{This figure presents KDE plots showing the distribution of differences in predicted values between modified and unmodified graphs. Each line within the plots corresponds to a distinct edge sampling strategy: sampling by affinity (blue), sampling by wealth (red), sampling by importance (green), and uniform sampling (yellow). Means are plotted in black.}
    \label{fig:updating_edges}
\end{figure}

Adding important edges increased predicted percentile floor price on average when compared to the unmodified graph, and this effect became more pronounced as more edges were added. When we added $100$ edges, for example, the GNN predicted a higher percentile floor price $99.86\%$ of the time. \Cref{subfig:adding_edges} shows the distribution of changes in predicted percentile floor price when adding edges. Note that when sampling by importance (plotted in green), the  floor price the model predicted increased by an average of $1.253$ percentiles. We observed similar trends when deleting edges; \cref{subfig:deleting_edges} shows the associated distribution of changes. These results were robust across the different numbers of edges added and deleted.\footnote{Results for all sample weightings and numbers of edges added or deleted are presented in \cref{app:band_tbls}.}

We next performed an ablation of both affinity and wealth. We observed that when adding edges to wallets sampled by their affinity, the GNN predicted a $56.9\%$ higher floor price on average on the modified graph, relative to when sampling by wealth; moreover, sampling by affinity was associated with higher value predictions overall. However, wealth still provided a meaningful signal, as sampling by importance outperformed affinity.  In \cref{subfig:adding_edges}, we plot the distribution of predicted differences in floor price when adding edges to wallets sampled by affinity (blue), by wealth (red), and uniformly (yellow) and their means. We also note that as more edges were added to the graph, the gap between the impact of affinity and wealth grew (see \cref{app:band_tbls}); this suggests that adding wallets from tightly connected communities may have a compounding effect on a collection's value.

Conversely, when edges were deleted from the graph, the GNN predicted lower values on average when sampling by wealth than by affinity. Furthermore, unlike when we added edges, the combined sampling approach offered fewer gains; importance changed the model's prediction by only $7.15\%$ more than the next best alternative, as compared to a $22.54\%$ difference when adding edges. This was likely because wallet affinity had less variance when restricted to owners of a specific collection, and was therefore a less informative signal of value. For the full results of predicted floor price movement when deleting edges, see \cref{subfig:deleting_edges}.

In addition, we performed an experiment in which we removed the bottom $n$th percentile of wallets by wealth per collection. Specifically, we tested this removal for every 5th percentile (i.e., 5th, 10th, 15th, etc.), and each time we observed that, on average, the predicted floor price of the affected collections decreased. This decline was statistically significant for each of the tested percentiles. These findings suggest that the network effect in the ownership graph extends beyond the most important or wealthiest wallets---having a broad and diverse community of wallet holders contributes meaningfully to the overall predicted value of an NFT collection. See \cref{tbl:wealth_percentile} in the appendix for the full table of results.

\section{The Snob Effect} \label{sec:snob}

In this section, we validate that NFT values are in part influenced by the snob effect.
We thus explore whether there is a negative correlation between the rarity rank of an NFT and its value within a collection, represented by its average sale price.\footnote{Negative correlation between sale price and a different notion of rarity was already demonstrated by \citet{mekacher2022heterogeneous} on a dataset of $410$ collections; however, at the time rarity rank was not easily viewable on marketplaces such as Opensea.} We expect the correlation to be negative because the \emph{rarest} NFT in a collection has the \emph{lowest} rarity rank.

\begin{figure*}
    \centering
  \begin{subfigure}[c]{0.4\textwidth}
    \centering%
    \includegraphics[width=\textwidth]{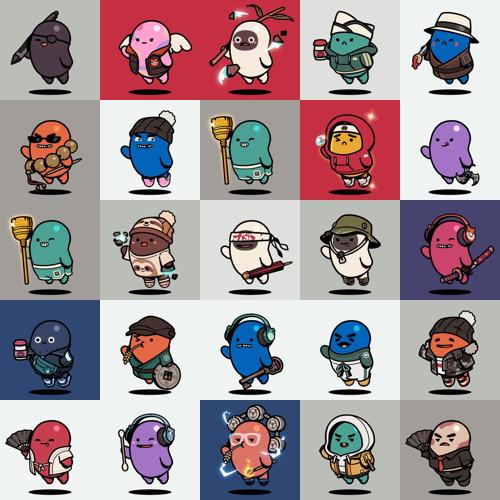}
    \caption{\capcapcap{Randomly selected Beanz NFTs.}}
    \label{fig:bean_grid}
  \end{subfigure}%
  \hspace{2cm}
  \begin{subfigure}[c]{0.45\textwidth}
    \centering%
    \begin{minipage}{0.48\textwidth}
      \centering%
      \includegraphics[width=\textwidth]{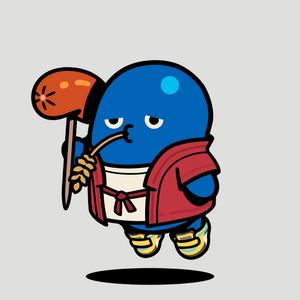}
    \end{minipage}
    \begin{minipage}{0.48\textwidth}
      \centering%
      \includegraphics[width=\textwidth]{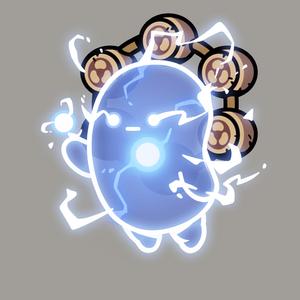}
    \end{minipage}
    \caption{\capcapcap{Left: Bean \#9848 -- most visually average in our sample. Right: Bean \#13956 -- most visually distinctive in our sample. }}
      \label{fig:rare_bean}
      \label{fig:com_bean}
  \end{subfigure}
  \caption{\capcap{The images with the least and greatest Euclidean distance to the centroid of the images in our small datasets subsample of the Beanz Originals collection.} Bean \#9848 [most average] had an average sale price of $1.6$ ETH ($3{,}700$ USD) across the sample period, while Bean \#13956 [most distinctive] had an average sale price of $40.7$ ETH ($93{,}700$ USD). Images \copyright~Azuki Labs, Inc.; used with permission.} 
  \Description[The images with the least and greatest Euclidean distance to the centroid of the images in our small datasets subsample of the Beanz Originals collection.]{Bean \#9848 [most average] had an average sale price of $1.6$ ETH ($3{,}700$ USD) across the sample period, while Bean \#13956 [most distinctive] had an average sale price of $40.7$ ETH ($93{,}700$ USD).}
  \label{fig:beanz}
\end{figure*}

As mentioned in \cref{sec:dataset}, rarity ranks are not available for all NFTs. Frequently, these are NFTs without randomly generated traits, making it difficult to calculate an explicit quantitative rarity ranking. We thus rely instead on a quantification of each NFT's \emph{visual distinctiveness}, which we define as the Euclidean distance between its embedding and the centroid embedding of its collection. This distance represents how distinct each NFT is from the ``average'' NFT in a collection; for example, the images with the smallest and greatest visual distinctiveness within the sample of the Beanz Originals collection in our dataset are pictured in \cref{fig:beanz}. While NFT buyers clearly do not compute these representations directly, they  likely do qualitatively assess visual distinctiveness.  We aim to determine whether, within an NFT collection, an NFT's visual distinctiveness positively correlates with its average sale price. We also compare the relative power of visual distinctiveness and rarity rank as predictors of value because understanding when one is more important than the other may provide insight for conspicuous goods markets beyond NFTs. 
We then show that the value increase from rarity is complementary with the value increase from popularity by showing it has a multiplicative effect on average collection value. Finally, we end the section with a case study of $9$ top-sales-volume collections in which we take a deeper look into how these measures impact the sales price and number of transaction for each token in each collection. 

\subsection{Evidence of the Snob Effect}\label{subsubsec:testing_snob}

We begin by examining the relationship between rarity rank and average sale price. We draw these values from our smaller dataset. We calculate the Pearson correlation between rarity rank and average sale price for each collection. 

It is worth noting that the NFT market tends to be relatively illiquid at the level of individual NFTs and quite volatile over long time scales. Consequently, two NFTs that most consumers value similarly might nevertheless exhibit significant differences in average historical sale prices, introducing noise into the sales data. 
Even so, of the NFT collections with rarity ranks available, $67.6\%$ exhibited statistically significant ($p<0.05$) negative correlation between rarity rank and average sale price. Conversely, only $1.0\%$ of collections exhibited significant positive correlation. We repeated this analysis restricted to only PFP collections; since users identify themselves with their profile pictures, we expected a more pronounced snob effect. However, we only observed a small change, with $70.9\%$ of PFP collections showing significant negative correlation between rarity rank and price and $1.0\%$ showing significant positive correlation. The full results can be found in \cref{table:snob_correlation}.

We now examine the relationship between visual distinctiveness and average NFT sale price. We considered both the entire dataset and only collections for which rarity ranks were not available. We computed the Pearson correlation between the visual distinctiveness and average sale price for each collection.

\begin{table*}[ht]
    \centering
    \begin{tabular}{lC{2.5cm}ccccc}
        \toprule
        & Category & Corr.~$(+)$ & Corr.~$(-)$ & \# Collections & Percent $(-/+)$ \\
        Predictor & & & & &\\
        \midrule
        \multirow{2}{*}{\parbox{2.2cm}{\centering Rarity Ranks}}& w/ rarity ranks & $10$ & $648$ & $959$ & $67.5\%$ $(-)$ \\ 
        & PFPs & $7$ & $462$ & $651$ & $70.9\%$ $(-)$ \\ 
        & non-PFPs & $3$ & $186$ & $308$ & $60.4\%$ $(-)$ \\ 
        \cmidrule{2-6}
        \multirow{4}{*}{\parbox{2.2cm}{\centering Visual Distinctiveness}}
        & All & $475$ & $18$ & $1265$ & $37.5\%$ $(+)$ \\ 
        & w/ rarity ranks & $392$ & $12$ & $959$ & $40.8\%$ $(+)$ \\ 
        & w/o rarity ranks & $81$ & $3$ & $326$ & $24.8\%$ $(+)$ \\
        & PFPs  & $313$ & $7$ & $760$ & $41.2\%$ $(+)$ \\ 
        & non-PFPs  & $162$ & $11$ & $505$ & $32.1\%$ $(+)$ \\ 
        \bottomrule
    \end{tabular}
    \caption{\capcap{Comparison of rarity rank and visual distinctiveness as predictors of average sale price across different NFT categories.} This table presents, for each category of NFTs, the number of NFT collections that have significant negative correlation with rarity rank and significant positive correlation with visual distinctiveness $(p<0.05)$.}
    \label{table:snob_correlation}
\end{table*}

We observed that $24.8\%$ of NFT collections without rarity ranks exhibited significant ($p<0.05$) positive correlation between visual distinctiveness and average sale price, with only $0.9\%$ collections showing significant negative correlation.  When examining all collections in our small dataset including those with rarity ranks, we observed $39.4\%$ showing significant positive correlation and only $0.1\%$ showing significant negative correlation. 
Our findings suggest that although visual distinctiveness is associated with average sale price within a collection, this association is not as strong as that of explicit rarity ranks when they are available. One possible explanation is that NFT marketplaces often provide features to easily sort and filter NFTs by rarity, making it straightforward for collectors to assess and trade on rarity, whereas visual distinctiveness is a more subjective and ad hoc measure. We investigated whether rarity ranks always account for more variance in sale prices compared to visual distinctiveness, or if there are collections where the trend is reversed. To determine which factor explains more variance in price, we fit two univariate linear regression models for each of the $959$ collections for which we had rarity ranks: predicting  price based on rarity rank and based on visual distinctiveness. We then compared the $R$-squared values of the two models for each collection, excluding those where neither model showed a positive $R$-squared. Among the $912$ remaining collections, rarity ranks explained more variance $71.5\%$ of the time, while visual distinctiveness was more predictive $28.5\%$ of the time. Despite the fact that rarity tends to be a better predictor in most cases, there are settings in which visual distinctiveness appears to better explain sale price.

\subsection{Beyond Linear Correlation}

We briefly explored why visual distinctiveness might appear to be a better predictor in some settings. One potential explanation is that rarity ranks may be challenging for linear models. For example, there can be a massive (in particular, nonlinear) price gap between two adjacently ranked NFTs if one of them is a unique one-of-one. We therefore compare the two predictors under Spearman's rank correlation coefficient, which measures monotonic relationships (linear or not). 

We computed the Spearman coefficients of the $260$ collections where visual distinctiveness explained more variance than rarity ranks in sales data. We then excluded collections without a significant Spearman correlation $(p>0.05)$ for either predictor, narrowing down to $166$ collections. Within this subset, $90$ collections showed a higher Spearman coefficient for rarity ranks, while $66$ demonstrated a greater Spearman coefficient for visual distinctiveness. This suggests that a sizeable portion of the cases where visual distinctiveness explained more variance was likely due to a non-linear relationship between sale price and rarity.

One rationale for collections whose price is better explained by visual appearance, even in a non-linear model, is that rarity ranks may lose precision in distinguishing the most unique NFTs. It is not uncommon to have a group of NFTs that have a one-of-one (or otherwise especially distinctive) trait, and yet will be sorted by their more common traits when ranked by OpenRarity. In this event, the OpenRarity score may not reliably reflect their ``rarity'' as perceived by a prospective owner, whereas ranking by visual distinctiveness may provide a more accurate representation. Another explanation is simply that there are collections in which the visual appeal of an NFT image is especially important---for example, if that image is primarily being used as a digital avatar on a social media platform or in an online game. In such cases standing out (or being visually distinct) could have a lot of value. 

\subsection{Evidence for Complementarity}

In this section, we examine whether there is complementarity between demand for a collection and demand for rarer NFTs within that collection.  More precisely, we analyze how the average collection value is related to the relationship between rarity and NFT sale price.\footnote{We use the average sale price here because our observations span a longer time period, making it more stable than the floor price. In \cref{sec:bandwagon}, we used the floor price because it provides an instantaneous measure of a collection's market value at the time of data collection.} We compare the fit of three  alternative models: one where the effect is multiplicative, one where the effect is additive, and one with no effect. Specifically, we normalized each NFT's rarity rank by converting it into a quantile and then fitted linear models to all NFT sale prices. We fitted three such models, each controlling for the collection's fixed effect in a different way: (1) normalizing the sale price by dividing by the mean (multiplicative), (2) normalizing the sale price by subtracting the mean (additive), and (3) leaving the sale price unnormalized. In all three cases, we discarded outlier collections having average sale prices $< 0.0000001$ ETH.

Both the unnormalized model and the additive model showed no significant relationship between rarity and the normalized average sale price. However, the multiplicative model showed a significant relationship ($R^2 = 0.08$, $p = 0$), suggesting a multiplicative interaction between rarity and average collection value---i.e., reflecting complementarity between demand for a collection  and demand for rarity within that collection. 

\subsection{Case Studies}
\label{varsec:case_study}
To further explore the relationship between distance, rarity and NFT sale data, we investigated the entire transaction and image dataset we gathered for $9$ top collections as described in \cref{subsubsec:case_study_data}. We selected the collections for these case studies by starting with the $30$ collections with the highest total sales volume in our dataset and removing those collections that did not have rarity ranks and that our previous study ruled out for not having significant correlations between either sale price and rarity or sale price and visual distance. This left us with $9$ collections.\footnote{The initial filtering step actually left us with $10$ collections. However, transaction data from the Meebits collection appeared to exhibit substantial amounts of wash trading (self-trading intended to manipulate the price record), so we removed this collection from our analysis.} The case study dataset has two main advantages: first, it contains much more sale price data for every token, allowing for more statistical power; and second, it contains additional data on the number of times each token was sold, allowing for another dimension of analysis.

\begin{figure}
    \centering
    \begin{subfigure}{0.4\textwidth}
        \includegraphics[trim={1.65cm 0.5cm 1.8cm 0.6cm},clip,width=\textwidth]{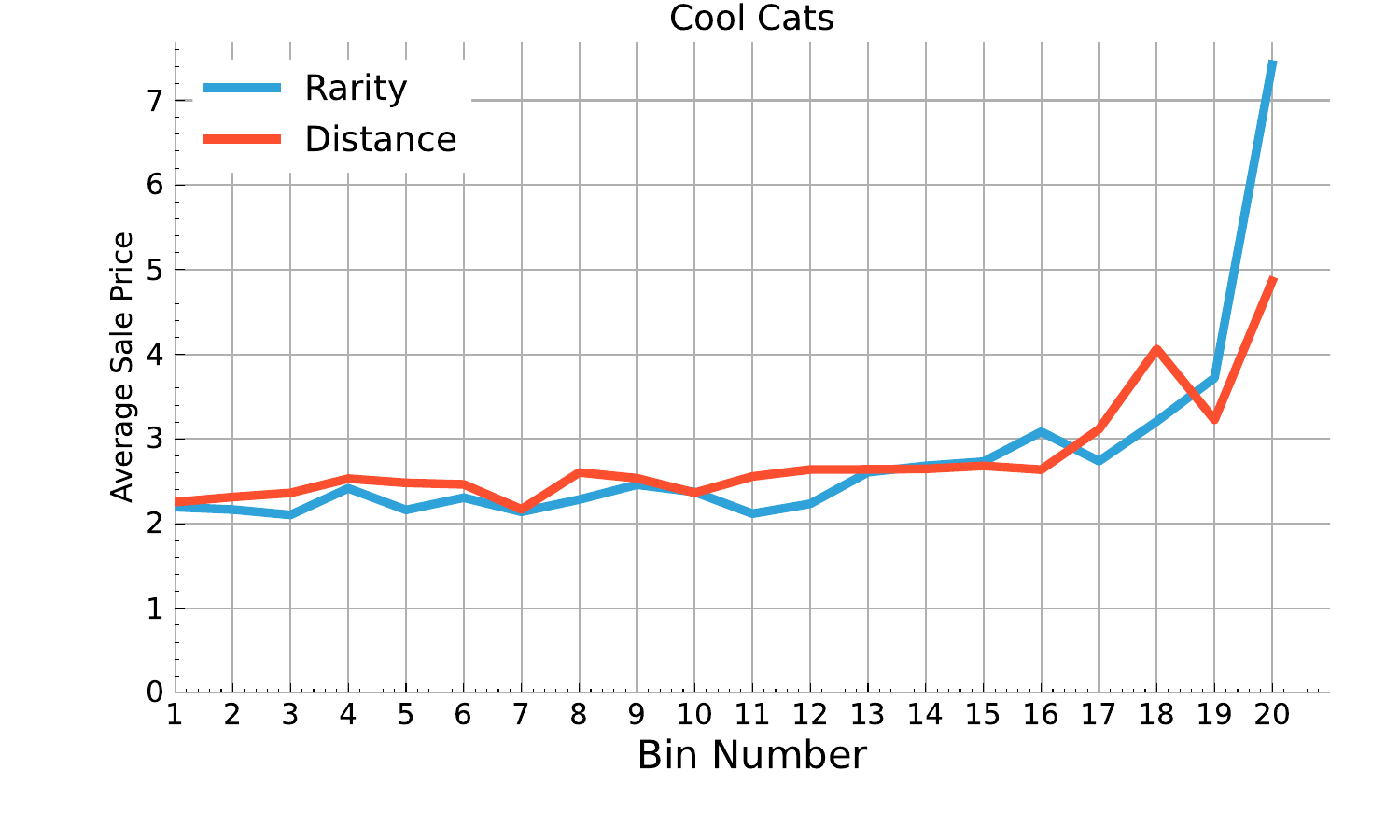}
        \subcaption{\capcapcap{Relationships with Sale Price}}
        \label{subfig:sale_price_quants}
    \end{subfigure}\hfill
    \begin{subfigure}{0.4\textwidth}
        \includegraphics[trim={1.8cm 0.45cm 1.8cm 0.6cm},clip,width=\textwidth]{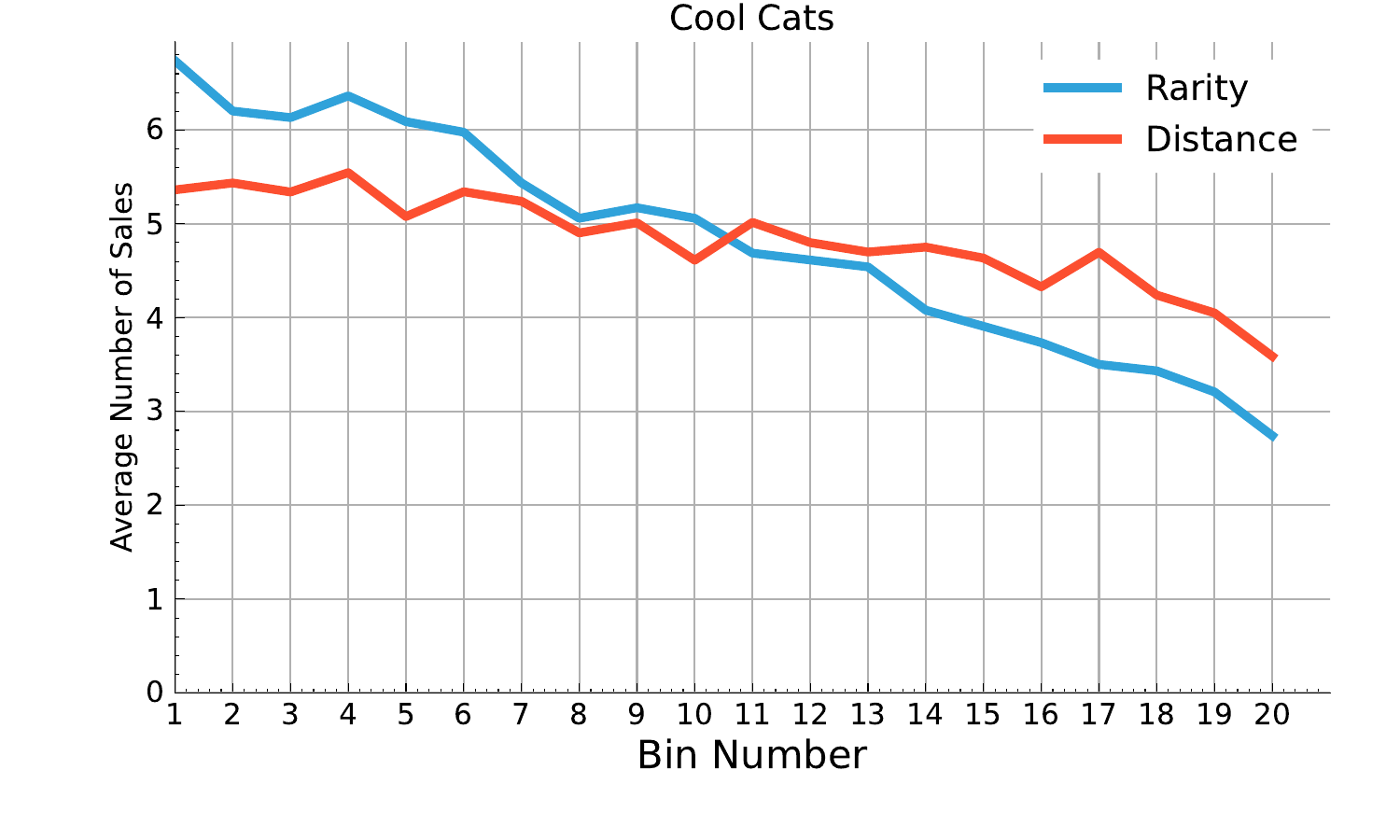}
        \subcaption{\capcapcap{Relationship with Number of Sales}}
        \label{subfig:num_sale_quants}
    \end{subfigure}
    \caption{\capcap{Relationship between quantile bins of rarity (visual distance) and sale price or number of sales respectively in the Cool Cats collection.} This figure presents rarity and visual distance placed into $20$ bins by quantiles such that each bin contains $5\%$ of the data. These bins are plotted against sale price and number of sales.  In the case of rarity ranks bins are sorted from highest rarity rank (least rare) to lowest rarity rank (most rare).
}
    \label{fig:case_study_plots}
\end{figure}

We began by replicating the qualitative analysis conducted by \citet{mekacher2022heterogeneous} on our dataset. 
\citeauthor{mekacher2022heterogeneous} analyzed $3$ ``exemplary'' collections by first binning the rarity of each collection's NFTs into $20$ quantiles; they observed that sale price was relatively flat in the lower quantiles but sharply increased in the last (most rare) $2$-$3$ buckets. We saw the same trend as \citet{mekacher2022heterogeneous} in each of the $9$ collections that we analyzed; additionally, in cases where visual distance and sale price were meaningfully correlated, we saw a similar relationship, albeit much less pronounced. \citet{mekacher2022heterogeneous} also analyzed the relationship between rarity and number of sales and found a positive relationship. We also observed that the relationship between rarity rank and number of sales appears to have been less driven by outlier values than the relationship between rarity rank and sale price. However, for visual distance, we observed that the relationship with number of sales was usually small or non-existent. For an example of these relationships see \cref{fig:case_study_plots} (for full plots for each of the $9$ collections appear in \cref{app:case_study_figs}).

We also performed a quantitative analysis of correlations, paralleling our analysis in \cref{subsubsec:testing_snob}. In most cases, distance and rarity rank had similar effects on each collection in terms of sale price: typically either both having a negligible effect (Pearson coefficient around $0.05$) or both having a more substantial effect (Pearson coefficient greater than $0.1$).\footnote{We discuss magnitude rather than $p$-value in this section because we have enough to data for almost every relationship to be statistically significant.} There are two collections where visual distance and rarity differed in their ability to explain sale price---Azuki and Mutant Ape Yacht Club---yet both of these exceptions in some sense prove the rule. In the case of Azuki, the most visually distinctive NFT images depict the character holding a boombox, which is neither particularly rare nor valuable relative to the rest of the collection. However, they are quite different from the samurai-esque aesthetic of most other images in the collection have. Similarly, in the case of Mutant Ape Yacht Club, the most visually distinctive NFTs depict the apes covered in worms rather than clothes, which is not a particularly valued trait in the community. These results suggest that a more nuanced notion of visual distance, one that allows for multiple clusters within a collection to deal with the existence of multiple common aesthetics, could be important for further analysis.

Next, we quantitatively measured how much of the relationship between rarity (resp.~visual distinctiveness) and sale price was driven by the most rare or visually distinctive NFTs. To do this, we recomputed correlation with the last $2$ buckets ($10$ percentiles) censored and examined how the correlation coefficient changed. In both cases, rarity and visual distance, we saw a relatively large drop in correlation coefficients with only~$1$ collection having a Pearson coefficient greater than $0.1$ for distance and only~$2$ collections having a Pearson coefficient greater than $0.1$ for rarity. This adds support for the idea that the relationship between sale price and rarity (resp.~visual distance) was driven by the most rare (resp.~visually distinctive) while the less rare (resp.~visually distinctive) almost shared an equivalence class.

Finally, we measured the correlation between rarity (resp.~visual distance) and number of sales, and found that rarity tended to be even more correlated with number of sales than with sale price. In the case of Cool Cats, for example, the Pearson coefficient jumped from $0.14$ to $0.28$. This is potentially unsurprising because the number of sales is a cumulative measure and therefore exhibits much less noise than average sale price, which is confounded by changes in the overall market. However, potentially more surprising is that visual distance tended to be less correlated with number of sales than with sale price. One potential reason for this is that visually distinctive NFTs include both the interesting and more desirable NFTs in a collection but also the ``uglier'' and more frequently turned over NFTs (e.g., the worm-coated Mutant Apes), creating a lot of variation. (For a full table of the Pearson coefficients described in this section, see \cref{app:case_study_tbls}.)

\section{Conclusion}\label{sec:conc}

This paper introduced a model of conspicuous consumption that shows that the bandwagon effect and snob effect  can coexist while interacting at different levels of resolution. It also validated this model by leveraging publicly available NFT data. The richness of our data also allowed us to examine the structure of social relationships within a conspicuous goods market for what, to our knowledge, is the first time.

In our analysis of bandwagon effects, we found that simulating an increase in the number of owners with high affinity for an NFT collection significantly increased that NFT collection's predicted value. The predictive strength of affinity in our trained model suggests that tight community structures may be important drivers of NFT value, a finding that is consistent with anecdotal and ethnographic accounts (see, e.g., \cite{casale2022impact, kaczynski2023everything, ante2024polychotomy}). Our analysis of the snob effect showed that publicly available rarity ranks tended to explain more variance in price than visual dissimilarity, suggesting that signals of exclusivity that are more easily understood and internalized by the market may be especially relevant for determining value. It will be important for future research to investigate the extent to which these patterns extend to other conspicuous goods markets.

It could be fruitful to consider refinements of the image embedding techniques used in this paper. As mentioned in \cref{varsec:case_study}, some NFT collections contain multiple distinct aesthetics; in such cases, centroids lose information by averaging these aesthetics together. One potential solution would be to allow for multiple clusters within a collection, each with its own centroid; we could then measure each NFT's distance from the nearest centroid as a more accurate measure of intra-collection distances. Another potentially useful refinement is fine-tuning the vision transformer on NFT images with the task of classifying the collection an NFT belongs to; this might produce image embeddings that more tightly cluster NFTs in the same collection.

Considering topics for future work more broadly, we note that this paper only scratches the surface of potential uses of NFT data for studying conspicuous goods. 
There are many additional questions about conspicuous consumption that have so far mainly been studied without access to purchase data. For example, qualitative research has looked at the effect of different types of scarcity on consumers' assessment of the value of conspicuous goods, suggesting that consumers' value assessments may be sensitive to supply-side scarcity (``limited edition'') but not sensitive to demand-side scarcity (``almost sold out'') \citep{gierl2010scarce}. By looking at the impact of collection size (which impacts supply-side scarcity) and market liquidity (which impacts demand-side scarcity) on NFT value, it might be possible to shed light on the magnitude of this effect---although of course there are challenges here because demand-side scarcity may also be a direct proxy for value in steady state. Another question worth investigating is about the existence of ``inconspicuous buyers,'' individuals who only care about signaling to their immediate peers and prefer signals that are hard to decipher by the broader ``masses''  \citep{makkar2018emotional, han2010signaling}. ``Inconspicuous consumption'' is effectively an inversion of the bandwagon effect, that has been thought to occur at the high end of the wealth distribution \citep{han2010signaling} or even just in well-connected individuals \cite{carbajal2015inconspicuous}.\footnote{That said, inconspicuous buying may be especially difficult in the NFT market, where the provenance of each asset can be directly traced, and wallet owners can often be re-identified even when they are nominally anonymous or pseudonymous.}
Given that we have access to the full contents of NFT owners' digital wallets, it may be possible to measure how conspicuous consumption dynamics vary with wealth level, or across different subcommunities of the market.

Despite its usefulness as a test bed for studying conspicuous consumption, there do exist senses in which the NFT market is different from other conspicuous goods markets. For example, the NFT market is currently patronized by a narrow band of consumers who skew tech-savvy and higher-income (see the discussion in \cite{kaczynski2023everything}). Conspicuous goods exist across the full range of social strata, and their dynamics may vary across consumer demographics.

Finally, we note that, while data on physical luxury markets is currently hard to come by, a growing wave of NFTs with associated physical-good counterparts (see, e.g., \cite{hasan2023using,natalee2023pairing,kaczynski2023everything}) suggests that it may one day be possible to conduct a version of our analysis for a class of physical goods as well. 
\section*{Acknowledgments and Disclosures}
{\footnotesize
Thanks to Kate Dellolio, Gerald Ferguson, Steve Kaczynski, Tim Roughgarden, Tynan Seltzer, Seraph, and Elena Silenok for helpful conversations. 

Lundy, Raman, and Leyton-Brown acknowledge support from an NSERC Discovery Grant, a CIFAR Canada AI Research Chair (Alberta Machine Intelligence Institute), and computational resources provided both by UBC Advanced Research Computing and a Digital Research Alliance of Canada RAC Allocation.

Kominers acknowledges support from the Digital Data Design ($\text{D}^3$) Institute at Harvard and the Ng Fund and the Mathematics in Economics Research Fund of the Harvard Center of Mathematical Sciences and Applications. 
Part of this work was conducted during the Simons Laufer Mathematical Sciences Institute Fall 2023 program on the Mathematics and Computer Science of Market and Mechanism Design, which was supported by the National Science Foundation under Grant No.~DMS-1928930 and by the Alfred~P.\ Sloan Foundation under grant G-2021-16778. 
Kominers is a Research Partner at a16z crypto, which reviewed a draft of this article for compliance prior to publication, and is an investor in crypto projects, including NFT projects (for general a16z disclosures, see \url{https://www.a16z.com/disclosures/}). Notwithstanding, the ideas and opinions expressed herein are those of the authors, rather than of a16z or its affiliates. 
Kominers also holds digital assets, including a small number of NFTs from some of the collections mentioned in this article; he also advises companies on web3 marketplace and incentive design, and serves as an expert on related matters.
}

\bibliographystyle{ACM-Reference-Format}

\bibliography{ref}


\begin{thebibliography}{33}


\ifx \showCODEN    \undefined \def \showCODEN     #1{\unskip}     \fi
\ifx \showDOI      \undefined \def \showDOI       #1{#1}\fi
\ifx \showISBNx    \undefined \def \showISBNx     #1{\unskip}     \fi
\ifx \showISBNxiii \undefined \def \showISBNxiii  #1{\unskip}     \fi
\ifx \showISSN     \undefined \def \showISSN      #1{\unskip}     \fi
\ifx \showLCCN     \undefined \def \showLCCN      #1{\unskip}     \fi
\ifx \shownote     \undefined \def \shownote      #1{#1}          \fi
\ifx \showarticletitle \undefined \def \showarticletitle #1{#1}   \fi
\ifx \showURL      \undefined \def \showURL       {\relax}        \fi
\providecommand\bibfield[2]{#2}
\providecommand\bibinfo[2]{#2}
\providecommand\natexlab[1]{#1}
\providecommand\showeprint[2][]{arXiv:#2}

\bibitem[Amaldoss and Jain(2008)]%
        {amaldoss2008research}
\bibfield{author}{\bibinfo{person}{Wilfred Amaldoss} {and} \bibinfo{person}{Sanjay Jain}.} \bibinfo{year}{2008}\natexlab{}.
\newblock \showarticletitle{Research note---Trading up: A strategic analysis of reference group effects}.
\newblock \bibinfo{journal}{\emph{Marketing Science}} \bibinfo{volume}{27}, \bibinfo{number}{5} (\bibinfo{year}{2008}), \bibinfo{pages}{932--942}.
\newblock


\bibitem[Ante(2024)]%
        {ante2024polychotomy}
\bibfield{author}{\bibinfo{person}{Lennart Ante}.} \bibinfo{year}{2024}\natexlab{}.
\newblock \showarticletitle{The Polychotomy of {NFT} ownership: Motivational heterogeneity and underlying drivers}.
\newblock \bibinfo{journal}{\emph{Digital Business}} (\bibinfo{year}{2024}), \bibinfo{pages}{100091}.
\newblock


\bibitem[Bekir et~al\mbox{.}(2013)]%
        {bekir2013luxury}
\bibfield{author}{\bibinfo{person}{Insaf Bekir}, \bibinfo{person}{Sana El~Harbi}, {and} \bibinfo{person}{Gilles Grolleau}.} \bibinfo{year}{2013}\natexlab{}.
\newblock \showarticletitle{How a luxury monopolist might benefit from the aspirational utility effect of counterfeiting?}
\newblock \bibinfo{journal}{\emph{European Journal of Law and Economics}}  \bibinfo{volume}{36} (\bibinfo{year}{2013}), \bibinfo{pages}{169--182}.
\newblock


\bibitem[Belk(1988)]%
        {belk1988possessions}
\bibfield{author}{\bibinfo{person}{Russell~W Belk}.} \bibinfo{year}{1988}\natexlab{}.
\newblock \showarticletitle{Possessions and the extended self}.
\newblock \bibinfo{journal}{\emph{Journal of Consumer Research}} \bibinfo{volume}{15}, \bibinfo{number}{2} (\bibinfo{year}{1988}), \bibinfo{pages}{139--168}.
\newblock


\bibitem[Carbajal et~al\mbox{.}(2015)]%
        {carbajal2015inconspicuous}
\bibfield{author}{\bibinfo{person}{Juan~Carlos Carbajal}, \bibinfo{person}{Jonathan Hall}, {and} \bibinfo{person}{Hongyi Li}.} \bibinfo{year}{2015}\natexlab{}.
\newblock \showarticletitle{Inconspicuous conspicuous consumption}.
\newblock \bibinfo{journal}{\emph{Working Paper}} (\bibinfo{year}{2015}).
\newblock


\bibitem[Casale-Brunet et~al\mbox{.}(2022)]%
        {casale2022impact}
\bibfield{author}{\bibinfo{person}{Simone Casale-Brunet}, \bibinfo{person}{Mirko Zichichi}, \bibinfo{person}{Lee Hutchinson}, \bibinfo{person}{Marco Mattavelli}, {and} \bibinfo{person}{Stefano Ferretti}.} \bibinfo{year}{2022}\natexlab{}.
\newblock \showarticletitle{The impact of {NFT} profile pictures within social network communities}. In \bibinfo{booktitle}{\emph{Proceedings of the 2022 ACM Conference on Information Technology for Social Good}}. \bibinfo{pages}{283--291}.
\newblock


\bibitem[Chaudhuri and Majumdar(2006)]%
        {chaudhuri2006diamonds}
\bibfield{author}{\bibinfo{person}{Himadri~Roy Chaudhuri} {and} \bibinfo{person}{Sitanath Majumdar}.} \bibinfo{year}{2006}\natexlab{}.
\newblock \showarticletitle{Of diamonds and desires: Understanding conspicuous consumption from a contemporary marketing perspective}.
\newblock \bibinfo{journal}{\emph{Academy of Marketing Science Review}}  \bibinfo{volume}{2006} (\bibinfo{year}{2006}), \bibinfo{pages}{1}.
\newblock


\bibitem[Chen et~al\mbox{.}(2016)]%
        {chen2016research}
\bibfield{author}{\bibinfo{person}{Li~Dan Chen}, \bibinfo{person}{Xin Lai}, \bibinfo{person}{Neng~Min Wang}, {and} \bibinfo{person}{Wei Huang}.} \bibinfo{year}{2016}\natexlab{}.
\newblock \showarticletitle{Research in Progress: the Snob and bandwagon effects on Consumers' Purchase Intention under Different Promotion Strategies}. In \bibinfo{booktitle}{\emph{{PACIS 2016 Proceedings}}}. \bibinfo{pages}{118}.
\newblock


\bibitem[Costa et~al\mbox{.}(2023)]%
        {costa2023show}
\bibfield{author}{\bibinfo{person}{Davide Costa}, \bibinfo{person}{Lucio La~Cava}, {and} \bibinfo{person}{Andrea Tagarelli}.} \bibinfo{year}{2023}\natexlab{}.
\newblock \showarticletitle{Show me your {NFT} and {I} tell you how it will perform: Multimodal representation learning for {NFT} selling price prediction}. In \bibinfo{booktitle}{\emph{Proceedings of the ACM Web Conference 2023}} (Austin, TX, USA) \emph{(\bibinfo{series}{WWW '23})}. \bibinfo{publisher}{Association for Computing Machinery}, \bibinfo{address}{New York, NY, USA}, \bibinfo{pages}{1875–1885}.
\newblock
\showISBNx{9781450394161}
\urldef\tempurl%
\url{https://doi.org/10.1145/3543507.3583520}
\showDOI{\tempurl}


\bibitem[{Dworczak, P. \textregistered\ M. Reuter \textregistered\ S. D. Kominers \textregistered\ C. Lee}(2024)]%
        {dworczak2024optimal}
\bibfield{author}{\bibinfo{person}{{Dworczak, P. \textregistered\ M. Reuter \textregistered\ S. D. Kominers \textregistered\ C. Lee}}.} \bibinfo{year}{2024}\natexlab{}.
\newblock \showarticletitle{Optimal Membership Design}.
\newblock \bibinfo{journal}{\emph{Working Paper}} (\bibinfo{year}{2024}).
\newblock


\bibitem[Gao(2023)]%
        {nakamigos2023}
\bibfield{author}{\bibinfo{person}{Jade Gao}.} \bibinfo{year}{April 3, 2023}\natexlab{}.
\newblock \showarticletitle{Decode the Rapid Success of {Nakamigos} Like A Pro}.
\newblock \bibinfo{journal}{\emph{Dapp Radar}} (\bibinfo{year}{April 3, 2023}).
\newblock


\bibitem[Geng and Chen(2019)]%
        {geng2019optimal}
\bibfield{author}{\bibinfo{person}{Wei Geng} {and} \bibinfo{person}{Zuguang Chen}.} \bibinfo{year}{2019}\natexlab{}.
\newblock \showarticletitle{Optimal pricing of virtual goods with conspicuous features in a freemium model}.
\newblock \bibinfo{journal}{\emph{International Journal of Electronic Commerce}} \bibinfo{volume}{23}, \bibinfo{number}{3} (\bibinfo{year}{2019}), \bibinfo{pages}{427--449}.
\newblock


\bibitem[Gierl and Huettl(2010)]%
        {gierl2010scarce}
\bibfield{author}{\bibinfo{person}{Heribert Gierl} {and} \bibinfo{person}{Verena Huettl}.} \bibinfo{year}{2010}\natexlab{}.
\newblock \showarticletitle{Are scarce products always more attractive? {The} interaction of different types of scarcity signals with products' suitability for conspicuous consumption}.
\newblock \bibinfo{journal}{\emph{International Journal of Research in Marketing}} \bibinfo{volume}{27}, \bibinfo{number}{3} (\bibinfo{year}{2010}), \bibinfo{pages}{225--235}.
\newblock


\bibitem[Goetz and Lu(2022)]%
        {goetz2022peer}
\bibfield{author}{\bibinfo{person}{Daniel Goetz} {and} \bibinfo{person}{Wei Lu}.} \bibinfo{year}{2022}\natexlab{}.
\newblock \showarticletitle{Peer Effects from Friends and Strangers: Evidence from Random Matchmaking in an Online Game}. In \bibinfo{booktitle}{\emph{Proceedings of the 23rd {ACM} Conference on Economics and Computation}}. \bibinfo{pages}{285--286}.
\newblock


\bibitem[Han et~al\mbox{.}(2010)]%
        {han2010signaling}
\bibfield{author}{\bibinfo{person}{Young~Jee Han}, \bibinfo{person}{Joseph~C Nunes}, {and} \bibinfo{person}{Xavier Dr{\`e}ze}.} \bibinfo{year}{2010}\natexlab{}.
\newblock \showarticletitle{Signaling status with luxury goods: The role of brand prominence}.
\newblock \bibinfo{journal}{\emph{Journal of Marketing}} \bibinfo{volume}{74}, \bibinfo{number}{4} (\bibinfo{year}{2010}), \bibinfo{pages}{15--30}.
\newblock


\bibitem[Hasan et~al\mbox{.}(2023)]%
        {hasan2023using}
\bibfield{author}{\bibinfo{person}{Haya~R. Hasan}, \bibinfo{person}{Mohammad Madine}, \bibinfo{person}{Ibrar Yaqoob}, \bibinfo{person}{Khaled Salah}, \bibinfo{person}{Raja Jayaraman}, {and} \bibinfo{person}{Dragan Boscovic}.} \bibinfo{year}{2023}\natexlab{}.
\newblock \showarticletitle{Using {NFTs} for ownership management of digital twins and for proof of delivery of their physical assets}.
\newblock \bibinfo{journal}{\emph{Future Generation Computer Systems}}  \bibinfo{volume}{146} (\bibinfo{year}{2023}), \bibinfo{pages}{1--17}.
\newblock


\bibitem[Hopkins(2024)]%
        {hopkins2024cardinal}
\bibfield{author}{\bibinfo{person}{Ed Hopkins}.} \bibinfo{year}{2024}\natexlab{}.
\newblock \showarticletitle{Cardinal sins? {Conspicuous} consumption, cardinal status and inequality}.
\newblock \bibinfo{journal}{\emph{Journal of the European Economic Association}} \bibinfo{volume}{22}, \bibinfo{number}{5} (\bibinfo{year}{2024}), \bibinfo{pages}{2374--2413}.
\newblock


\bibitem[Kaczynski and Kominers(2024)]%
        {kaczynski2023everything}
\bibfield{author}{\bibinfo{person}{Steve Kaczynski} {and} \bibinfo{person}{Scott~Duke Kominers}.} \bibinfo{year}{2024}\natexlab{}.
\newblock \bibinfo{booktitle}{\emph{The Everything Token: How {NFTs} and {Web3} Will Transform the Way We Buy, Sell, and Create}}.
\newblock \bibinfo{publisher}{Portfolio}.
\newblock


\bibitem[Ko et~al\mbox{.}(2022)]%
        {ko2022economic}
\bibfield{author}{\bibinfo{person}{Hyungjin Ko}, \bibinfo{person}{Bumho Son}, \bibinfo{person}{Yunyoung Lee}, \bibinfo{person}{Huisu Jang}, {and} \bibinfo{person}{Jaewook Lee}.} \bibinfo{year}{2022}\natexlab{}.
\newblock \showarticletitle{The economic value of {NFT}: Evidence from a portfolio analysis using mean--variance framework}.
\newblock \bibinfo{journal}{\emph{Finance Research Letters}}  \bibinfo{volume}{47} (\bibinfo{year}{2022}), \bibinfo{pages}{102784}.
\newblock


\bibitem[Kuwashima(2016)]%
        {kuwashima2016structural}
\bibfield{author}{\bibinfo{person}{Yufu Kuwashima}.} \bibinfo{year}{2016}\natexlab{}.
\newblock \showarticletitle{Structural equivalence and cohesion can explain bandwagon and snob effect}.
\newblock \bibinfo{journal}{\emph{Annals of Business Administrative Science}} \bibinfo{volume}{15}, \bibinfo{number}{1} (\bibinfo{year}{2016}), \bibinfo{pages}{1--14}.
\newblock


\bibitem[Leibenstein(1950)]%
        {leibenstein1950bandwagon}
\bibfield{author}{\bibinfo{person}{Harvey Leibenstein}.} \bibinfo{year}{1950}\natexlab{}.
\newblock \showarticletitle{Bandwagon, snob, and {Veblen} effects in the theory of consumers' demand}.
\newblock \bibinfo{journal}{\emph{Quarterly Journal of Economics}} \bibinfo{volume}{64}, \bibinfo{number}{2} (\bibinfo{year}{1950}), \bibinfo{pages}{183--207}.
\newblock


\bibitem[Lundy et~al\mbox{.}(2024)]%
        {lundy2024pay}
\bibfield{author}{\bibinfo{person}{Taylor Lundy}, \bibinfo{person}{Narun Raman}, \bibinfo{person}{Hu Fu}, {and} \bibinfo{person}{Kevin Leyton-Brown}.} \bibinfo{year}{2024}\natexlab{}.
\newblock \showarticletitle{Pay to (Not) Play: Monetizing Impatience in Mobile Games}. In \bibinfo{booktitle}{\emph{Proceedings of the AAAI Conference on Artificial Intelligence}}, Vol.~\bibinfo{volume}{38}. \bibinfo{pages}{9856--9864}.
\newblock


\bibitem[Makkar and Yap(2018)]%
        {makkar2018emotional}
\bibfield{author}{\bibinfo{person}{Marian Makkar} {and} \bibinfo{person}{Sheau-Fen Yap}.} \bibinfo{year}{2018}\natexlab{}.
\newblock \showarticletitle{Emotional experiences behind the pursuit of inconspicuous luxury}.
\newblock \bibinfo{journal}{\emph{Journal of Retailing and Consumer Services}}  \bibinfo{volume}{44} (\bibinfo{year}{2018}), \bibinfo{pages}{222--234}.
\newblock


\bibitem[Mekacher et~al\mbox{.}(2022)]%
        {mekacher2022heterogeneous}
\bibfield{author}{\bibinfo{person}{Amin Mekacher}, \bibinfo{person}{Alberto Bracci}, \bibinfo{person}{Matthieu Nadini}, \bibinfo{person}{Mauro Martino}, \bibinfo{person}{Laura Alessandretti}, \bibinfo{person}{Luca~Maria Aiello}, {and} \bibinfo{person}{Andrea Baronchelli}.} \bibinfo{year}{2022}\natexlab{}.
\newblock \showarticletitle{Heterogeneous rarity patterns drive price dynamics in {NFT} collections}.
\newblock \bibinfo{journal}{\emph{Scientific Reports}} \bibinfo{volume}{12}, \bibinfo{number}{1} (\bibinfo{year}{2022}), \bibinfo{pages}{13890}.
\newblock


\bibitem[Nadini et~al\mbox{.}(2021)]%
        {nadini2021mapping}
\bibfield{author}{\bibinfo{person}{Matthieu Nadini}, \bibinfo{person}{Laura Alessandretti}, \bibinfo{person}{Flavio Di~Giacinto}, \bibinfo{person}{Mauro Martino}, \bibinfo{person}{Luca~Maria Aiello}, {and} \bibinfo{person}{Andrea Baronchelli}.} \bibinfo{year}{2021}\natexlab{}.
\newblock \showarticletitle{Mapping the {NFT} revolution: Market trends, trade networks, and visual features}.
\newblock \bibinfo{journal}{\emph{Scientific Reports}} \bibinfo{volume}{11}, \bibinfo{number}{1} (\bibinfo{year}{2021}), \bibinfo{pages}{20902}.
\newblock


\bibitem[Natalee(2024)]%
        {natalee2023pairing}
\bibfield{author}{\bibinfo{person}{Natalee}.} \bibinfo{year}{accessed September 7, 2024}\natexlab{}.
\newblock \showarticletitle{Pairing {NFTs} with Physical Assets: A New Approach to Authenticity and Ownership}.
\newblock \bibinfo{journal}{\emph{NFT Culture}} (\bibinfo{year}{accessed September 7, 2024}).
\newblock


\bibitem[Oh et~al\mbox{.}(2023)]%
        {digitalveblen2022}
\bibfield{author}{\bibinfo{person}{Sebeom Oh}, \bibinfo{person}{Samuel Rosen}, {and} \bibinfo{person}{Anthony~Lee Zhang}.} \bibinfo{year}{2023}\natexlab{}.
\newblock \showarticletitle{Digital {Veblen} Goods}.
\newblock \bibinfo{journal}{\emph{Working Paper}} (\bibinfo{year}{2023}).
\newblock


\bibitem[Oquab et~al\mbox{.}(2024)]%
        {oquab2023dinov2}
\bibfield{author}{\bibinfo{person}{Maxime Oquab}, \bibinfo{person}{Timothée Darcet}, \bibinfo{person}{Théo Moutakanni}, \bibinfo{person}{Huy Vo}, \bibinfo{person}{Marc Szafraniec}, \bibinfo{person}{Vasil Khalidov}, \bibinfo{person}{Pierre Fernandez}, \bibinfo{person}{Daniel Haziza}, \bibinfo{person}{Francisco Massa}, \bibinfo{person}{Alaaeldin El-Nouby}, \bibinfo{person}{Mahmoud Assran}, \bibinfo{person}{Nicolas Ballas}, \bibinfo{person}{Wojciech Galuba}, \bibinfo{person}{Russell Howes}, \bibinfo{person}{Po-Yao Huang}, \bibinfo{person}{Shang-Wen Li}, \bibinfo{person}{Ishan Misra}, \bibinfo{person}{Michael Rabbat}, \bibinfo{person}{Vasu Sharma}, \bibinfo{person}{Gabriel Synnaeve}, \bibinfo{person}{Hu Xu}, \bibinfo{person}{Hervé Jegou}, \bibinfo{person}{Julien Mairal}, \bibinfo{person}{Patrick Labatut}, \bibinfo{person}{Armand Joulin}, {and} \bibinfo{person}{Piotr Bojanowski}.} \bibinfo{year}{2024}\natexlab{}.
\newblock \showarticletitle{DINOv2: Learning Robust Visual Features without Supervision}.
\newblock \bibinfo{journal}{\emph{arXiv:2304.07193}} (\bibinfo{year}{2024}).
\newblock


\bibitem[Shimron(2022)]%
        {axie2022}
\bibfield{author}{\bibinfo{person}{Leeor Shimron}.} \bibinfo{year}{August 13, 2022}\natexlab{}.
\newblock \showarticletitle{{Axie Infinity}: Pernicious Pyramid Scheme Or Gaming Breakthrough?}
\newblock \bibinfo{journal}{\emph{Forbes}} (\bibinfo{year}{August 13, 2022}).
\newblock
\urldef\tempurl%
\url{https://www.forbes.com/sites/leeorshimron/2022/08/13/axie-infinity-pernicious-pyramid-scheme-or-gaming-breakthrough/}
\showURL{%
\tempurl}


\bibitem[Shukla(2008)]%
        {shukla2008conspicuous}
\bibfield{author}{\bibinfo{person}{Paurav Shukla}.} \bibinfo{year}{2008}\natexlab{}.
\newblock \showarticletitle{Conspicuous consumption among middle age consumers: Psychological and brand antecedents}.
\newblock \bibinfo{journal}{\emph{Journal of Product \& Brand Management}} \bibinfo{volume}{17}, \bibinfo{number}{1} (\bibinfo{year}{2008}), \bibinfo{pages}{25--36}.
\newblock


\bibitem[Thorstein(1899)]%
        {veblen1955}
\bibfield{author}{\bibinfo{person}{Veblen Thorstein}.} \bibinfo{year}{1899}\natexlab{}.
\newblock \bibinfo{booktitle}{\emph{The Theory of the Leisure Class}}.
\newblock \bibinfo{publisher}{New York: Mentor Books}.
\newblock


\bibitem[Vigneron and Johnson(1999)]%
        {vigneron1999review}
\bibfield{author}{\bibinfo{person}{Franck Vigneron} {and} \bibinfo{person}{Lester~W Johnson}.} \bibinfo{year}{1999}\natexlab{}.
\newblock \showarticletitle{A review and a conceptual framework of prestige-seeking consumer behavior}.
\newblock \bibinfo{journal}{\emph{Academy of Marketing Science Review}} \bibinfo{volume}{1}, \bibinfo{number}{1} (\bibinfo{year}{1999}), \bibinfo{pages}{1--15}.
\newblock


\bibitem[Wang et~al\mbox{.}(2023)]%
        {wang2023dissecting}
\bibfield{author}{\bibinfo{person}{Jying-Nan Wang}, \bibinfo{person}{Yen-Hsien Lee}, \bibinfo{person}{Hung-Chun Liu}, {and} \bibinfo{person}{Yuan-Teng Hsu}.} \bibinfo{year}{2023}\natexlab{}.
\newblock \showarticletitle{Dissecting returns of non-fungible tokens ({NFTs}): Evidence from {CryptoPunks}}.
\newblock \bibinfo{journal}{\emph{North American Journal of Economics and Finance}}  \bibinfo{volume}{65} (\bibinfo{year}{2023}), \bibinfo{pages}{101892}.
\newblock


\end{thebibliography}

\appendix
\clearpage

\section{Formal Model}\label{app:formal_model}

We frame our analysis with a simple model of conspicuous consumption incorporating both bandwagon and snob effects, building off of and extending the framework of \citet{hopkins2024cardinal}:\footnote{We use notation consistent with that of \citet{hopkins2024cardinal} where possible, although we switch Hopkins's sign convention for the parameter $\cob$.} A continuum of agents have income $z$ distributed according to $G(z)$, with non-zero density $g(z)$ on $[\zmin,\zmax]$. There are two goods, $x$ and $y$, with $x$ being conspicuous---i.e., its consumption is observed by others---and $y$ is a composite good representing other consumption (privately observed).

Under optimal consumption, no income is wasted; hence, at the optimum, we have $y=z-x$ for an agent with income $z$. Thus, agents' optimal strategies are entirely specified by their choice of conspicuous consumption level $x$; we assume these strategies are symmetric in income, leading to a (measurable) conspicuous consumption choice function $x(z)$ and aggregate conspicuous consumption distribution~$F(x)$.

Agents' utilities depend on both consumption levels and social interaction effects driven by conspicuous consumption: $$U=U(x,y,\soc),$$ where $\soc$ is the social interaction term. We assume that social interactions have two components, a network effect among consumers of the conspicuous good, and a status preference that depends on an agent's level of consumption of the conspicuous good relative to others in the population. Formally, we take $$\soc[\xnx]\equiv\coc\Nfn[\xnx]+(\cob\Afn[\xnx]-\coa\Dfn[\xnx]),$$ with $$\Afn[\xnx]=\int^{x}_{0}(x-t)dF(t)\equiv xF(x)-\afn[\xnx]$$ and $$\Dfn[\xnx]=\int_{x}^\infty(t-x)dF(t)\equiv \dfn[\xnx]-x(1-F(x))$$ respectively representing status advantage and disadvantage relative to the population. 

We assume that $\Nfn>0$, $1\geq\coc\geq 0$, and $1\geq\coa\geq\cob\geq0$, so that network benefits are positive and status is increasing in relative advantage but decreasing in relative disadvantage, with upwards comparisons being more significant for utility than downwards comparisons. 
We assume that $\Nfn$ is twice--continuously differentiable with (weakly) positive first derivatives and (weakly) positive cross-partial derivatives in all arguments, representing the idea that network benefits of conspicuous consumption for a given agent increase as other agents' investment in conspicuous consumption grows. We furthermore assume that $\Nfn$ is concave in each input.

Finally, we make a series of regularity assumptions on the utility function introduced by  \citet{hopkins2024cardinal}: (i) $U$ is twice--continuously differentiable; $U$ (ii) $U_x>0$, $U_y>0$, and $U_{\soc}>0$ (utility is increasing in consumption and total social interactions); (iii) $U_{xy}\geq 0$ and $U_{yx}\geq 0$ (conspicuous consumption and ordinary consumption are not substitutes); (iv) $U_{xx}<0$, $U_{yy}<0$, and $U_{\soc\soc}<0$ (concavity in individual inputs); (v) $U_{xy}-U_{yy}>0$ (strict normality); (vi) $U_x(x,z-x,\soc)-U_y(x,z-x,\soc)=0$ has a unique solution $\hat{x}$ (reflecting the privately optimal level of conspicuous consumption absent social interactions), and $$(U_{x\soc}(x,z-x,\soc)-(U_{y\soc}(x,z-x,\soc))(x-\hat{x})<0.\footnote{See \citet{hopkins2024cardinal} for discussion of the economic substance and implications of assumptions (v) and (vi).}$$

Now, in the case that $\coc=0$, Lemma 1 of \cite{hopkins2024cardinal} shows that own (optimal) consumption of the conspicuous good and others' consumption of the conspicuous good  are strategic complements whenever $F(x)$ is differentiable in own consumption. Given the direct complementarity between own and others' consumption in $\Nfn$, the strategic complementarity finding extends to the case of $\coc>0$ as well. For completeness and because it is useful in our subsequent analysis, we present a self-contained the argument here based on Hopkins's proof of his Lemma 1.

The first-order condition for agent $i$'s optimal choice of conspicuous consumption level $x$ is given by
\begin{equation}
U_x(x,z-x,\soc)-U_y(x,z-x,\soc)+ \left(\coc\Nfnx[\xnx]+(\cob F(x)+\coa(1-F(x)))\right)U_{\soc}(x,z-x,\soc)=0.
\label{eq:foc}
\end{equation}
 Under the hypothesis that $F(x)$ is differentiable, \eqref{eq:foc} defines an optimal $\optx$ as a function of $z$, $F(x)$, $\coc\Nfn[\xnx]$, and $\coa\dfn[\xnx]+\cob\afn[\xnx]$. Denoting the left side of \eqref{eq:foc} by $\psi$ and using the implicit function theorem gives:

\begin{align*}
\frac{\partial \optx}{\partial \dfn}&=-\frac{\frac{\partial\psi}{\partial\dfn}}{\frac{\partial\psi}{\partial x}}.
\end{align*}
Now, we have $$\frac{\partial\psi}{\partial\dfn}=\coa(-U_{\soc x}+U_{\soc y}-U_{\soc \soc}\soc_x).$$ We have $\soc_x=\coc\Nfn_x+\cob F(x)+\coa(1-F(x))>0$ (recalling that $\Nfn_x\geq 0$, $\coc\geq0$, $\cob\geq0$, and $\coa\geq0$); moreover we have $-U_{\soc x}+U_{\soc y}>0$ and $-U_{\soc \soc}\geq 0$ by assumptions (vi) and (iv), respectively. Thus, $\frac{\partial\psi}{\partial\dfn}>0$.

If $F(x)$ is differentiable at $x$, then we have
\begin{align}
\frac{\partial\psi}{\partial x}&= (U_{xx}-U_{xy} +U_{x\soc}\soc_x)
-(U_{yx}-U_{yy}+ U_{y\soc}\soc_x)\label{eq:deriv2b}
\\&\quad+(\coc\Nfnxx+(\cob-\coa)f(x))\,U_{\soc}\label{eq:deriv2c}
\\&\quad+(\coc\Nfnx+\cob F(x)+\coa(1-F(x)))\,U_{\soc\soc}\soc_x.\label{eq:deriv2d}
\end{align}
The right side of \eqref{eq:deriv2b} is negative by assumptions (iv), (v), and (vi); \eqref{eq:deriv2c} is negative by the concavity of $\Nfn$, the fact that $\coc>0$, the fact that $\cob\leq\coa$, and the fact that $U_{\soc}>0$; and \eqref{eq:deriv2d} is negative because $\Nfnx\geq 0$, $\coc>0$, $U_{\soc\soc}< 0$, and $\soc_x>0$.  

Combining the preceding observations, we see that $\frac{\partial\psi}{\partial\dfn}>0$. A similar argument shows that  $\frac{\partial\psi}{\partial\afn}>0$.

Finally, we examine \begin{align*}
\frac{\partial \optx}{\partial \Nfn}&=-\frac{\frac{\partial\psi}{\partial\Nfn}}{\frac{\partial\psi}{\partial x}}. \label{eq:derivN}
\end{align*} Again, $\frac{\partial\psi}{\partial x}<0$, so we just need to show that $\frac{\partial\psi}{\partial\Nfn}>0$; this follows argument analogous to that used above because $$\frac{\partial\psi}{\partial\Nfn}=\coc\Nfnx(-U_{\soc x}+U_{\soc y}-U_{\soc \soc}\soc_x)$$ and $\coc\Nfnx\geq 0$ by assumption. 

Note also that as $\nx$ grows, the contribution of $\frac{\partial\optx}{\partial\Nfn}$ to the total derivative of $\optx$ declines because the concavity of $\Nfn$ in (each element of) $\nx$ implies that $\Nfnx[\xnx]$ is declining in (each element of)  $\nx$. Meanwhile, $\coa$ and $\cob$ are fixed, so we see that the share of the  total derivative of $\optx$ reflected through the social comparison terms is growing on the margin relative to the network term. Thus, we see that increases from low levels of others' conspicuous consumption impact $\optx$ more through the network benefit component $\Nfn$ than 
through social comparison effects, relative to increases from larger levels of others' conspicuous consumption.

Our observations are summarized in the following proposition.

\begin{prop}\label{prop:1}
If $F(x)$ is differentiable in own consumption, then own consumption and others' consumption of the conspicuous good are strategic complements; indeed, we have $\frac{\partial \optx}{\partial \Nfn}>0$, $\frac{\partial \optx}{\partial \dfn}>0$, and $\frac{\partial \optx}{\partial \afn}>0$. Moreover, the relative share of the impact of the network term $\coc\Nfn$  (resp.~ the social comparison term $(\cob\Afn[\xnx]-\coa\Dfn[\xnx])$) on the degree of complementarity between own and others' conspicuous consumption is decreasing (resp.~increasing) in the level of others' conspicuous consumption.
\end{prop}

Finally, note also that the higher $\Nfnxj$, which we might think of as reflecting agent $j$ being particularly prominent in the network, in the sense that they generate a larger marginal network externality, is associated with a greater influence on the optimal conspicuous consumption of agents $i\neq j$.

\section{Embedding Stability} \label{app:embed}

\begin{figure}
    \centering
    \includegraphics[width=0.4\textwidth]{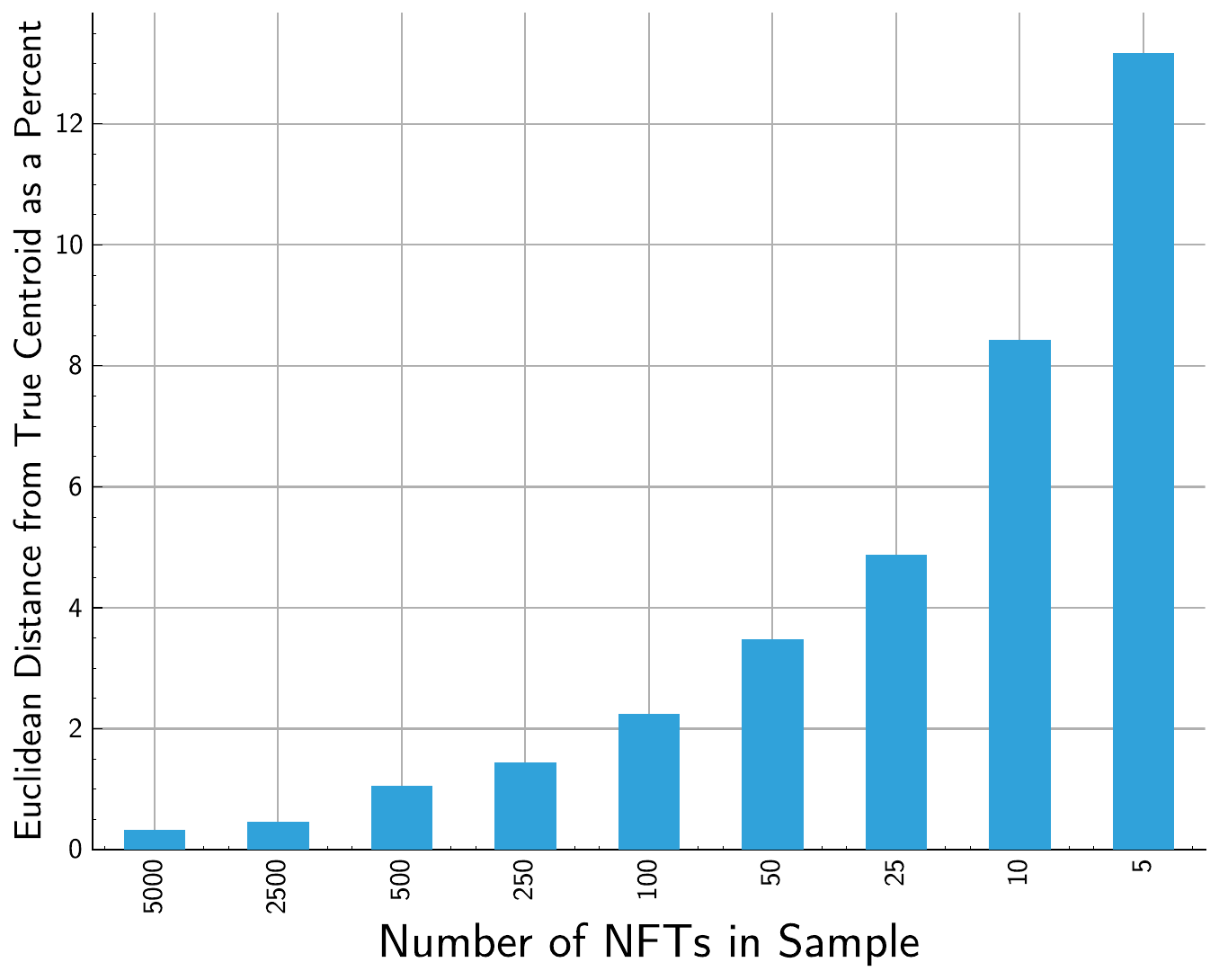} 
    \caption{\capcap{Average percent difference in distance to the ``true'' centroid and the subsample-constructed centroid.} For each subsample size $s$, we plot the average difference between an NFT's distance to the ``true'' centroid, computed using every NFT in the dataset, and its distance to the subsample centroid, computed only using a size-$s$ random subset of NFTs. The centroids are relatively stable (under $5\%$ change in average distance) as long as at least $25$ NFTs are used to compute them.}
    \Description[Average percent difference in distance to the ``true'' centroid and the subsample-constructed centroid.]{For each subsample size $s$, we plot the average difference between an NFT's distance to the ``true'' centroid, computed using every NFT in the dataset, and its distance to the subsample centroid, computed only using a size-$s$ random subset of NFTs. The centroids are relatively stable (under $5\%$ change in average distance) as long as at least $25$ NFTs are used to compute them.}
    \label{fig:stable}
\end{figure}

In order to evaluate how stable centroids are across various subsampling sizes, we created a dataset containing the full $5000$-$10000$ images for $10$ randomly-selected PFP NFT collections. (We rejection-sampled to ensure that the collections have at least $5000$ images.)  For each of those collections, we computed the ``true'' centroid that results from averaging all images and compared it to the centroid that is obtained by subsampling a smaller number of images. For each image in our dataset, we then compute distance to both the true centroid and the subsampled centroid, and averaged the absolute values of those differences. We found that even if only $50$ images were used to construct the centroid, on average the differences were only $4\%$ of the average distance to the true centroid. In \cref{fig:stable}, we plot the average percent difference across various sample sizes.

\section{Bandwagon Effect Experimental Setup}\label{app:bandwagon_exp_setup}

To predict floor price for PFP NFT collections, we utilized a Graph Neural Network (GNN) comprising four Graph Convolutional Network (GCN) layers. The architecture starts with an input layer accepting single-dimensional features (or $384$-dimensional features in the case of centroids), progressively transforming these through hidden layers with dimensions of $64$, $32$, and $16$, respectively---with an output layer that predicts a single-dimensional node feature. 

This experimental study was conducted on the Compute Canada computing cluster, leveraging both the Narval and Cedar resources as well as the Sockeye computing cluster. We used 40GB A100 GPUs on Narval, 32GB V100s on Cedar, and Dell EMC R440 CPUs on Sockeye. 

Due to limitations in GPU memory, we split the ownership graph into $50$ subgraphs. The splitting procedure was as follows: We first sampled $75$ wallets from the dataset, then we sampled all the collections they held (capping at $1500$), and then we pulled all the wallets that held those collections.

Because of our splitting procedure, we needed to be careful in how we split up our training, validation, and test sets. We sampled collection nodes for sets from the global list, and constructed a mapping into the subgraphs. We then trained our GNN with mean-squared error loss on a batch size of $4$ subgraphs for 2,500 total epochs. We took the model with the best validation accuracy across those 2,500 epochs which occurred at epoch 1,000.

\section{Supplementary Tables for Section \ref{sec:bandwagon}}\label{app:band_tbls}

\begin{table}[H]
    \centering
    \begin{tabular}{ccccc}
    \toprule
    & & Adding Edges & Deleting Edges \\
    Number of Edges & Sampling Procedure & & \\
    \midrule
    \multirow{4}{*}{25} & Affinity & $0.494$ & $-0.265$ \\
                         & Wealth & $0.448$ & $-0.531$ \\
                         & Importance & $0.575$ & $-0.612$ \\
                         & Uniform & $0.295$ & $-0.083$ \\
    \midrule
    \multirow{4}{*}{50} & Affinity & $0.714$ & $-0.348$ \\
                        & Wealth & $0.523$ & $-0.746$ \\
                        & Importance & $0.889$ & $-0.797$ \\
                        & Uniform & $0.468$ & $-0.336$ \\
    \midrule
    \multirow{4}{*}{100} & Affinity & $1.028$ & $-0.501$ \\
                        & Wealth & $0.655$ &  $-0.922$\\
                        & Importance & $1.253$ & $-0.988$ \\
                        & Uniform & $0.596$ & $-0.399$ \\
    \midrule
    \multirow{4}{*}{200} & Affinity & $1.481$ & $-0.614$ \\
                         & Wealth & $0.981$ & $-1.061$ \\
                         & Importance & $1.523$ & $-1.257$ \\
                         & Uniform & $0.798$ & $-0.474$ \\
    \bottomrule
    \end{tabular}
    \caption{\capcap{Average changes in value predictions.} This table shows the average change in the prediction of the floor price of all collections by the GNN across each of the graph modification settings. }
    \label{tbl:bandwagon}
\end{table}

\begin{table}[H]
    \centering
    \begin{tabular}{lr}
    \toprule
    Percentile &  Predicted Difference \\
    \midrule
    5          &   -6.507646 \\
    10         &   -6.609237 \\
    15         &   -6.788506 \\
    20         &   -7.010531 \\
    25         &   -7.371150 \\
    30         &   -7.804412 \\
    35         &   -8.240693 \\
    40         &   -8.695797 \\
    45         &   -9.044717 \\
    50         &   -9.320989 \\
    55         &   -9.533721 \\
    60         &   -9.640957 \\
    65         &   -9.663770 \\
    70         &   -9.688656 \\
    75         &   -9.645520 \\
    80         &   -9.538406 \\
    85         &   -9.270125 \\
    90         &   -8.619863 \\
    95         &   -7.166308 \\
    \bottomrule
    \end{tabular}
    \caption{\capcap{Average changes in value predictions when removing the edges to the bottom \emph{n}th percentile of wallets.} This table shows the average change in the prediction of the floor price of all collections by the GNN when removing the bottom \emph{n}th percentile of wallets by their wealth.}
    \label{tbl:wealth_percentile}
\end{table}

\section{Supplementary Tables for \Cref{varsec:case_study}}

\label{app:case_study_tbls}
\begin{table}[H]
\begin{tabular}{lrrr}
\toprule
Collection & Sale Price Corr. & \# of Sales Corr. & Censored Sale Price Corr. \\
\midrule
azuki & $0.083537$ & $-0.049221$ & $0.082964$ \\
beanzofficial & $0.126023$ & $-0.092573$ & $0.090254$ \\
boredapeyachtclub & $0.066870$ & $-0.051255$ & $0.037238$ \\
clonex & $0.130384$ & $-0.104769$ & $0.118309$ \\
cool-cats-nft & $0.128013$ & $-0.115599$ & $0.066445$ \\
doodles-official & $0.182818$ & $-0.122661$ & $0.075605$ \\
mutant-ape-yacht-club & $\emph{0.013094}$ & $0.039136$ & $\emph{0.003909}$ \\
proof-moonbirds & $0.183427$ & $-0.113823$ & $0.064193$ \\
pudgypenguins & $0.058748$ & $-0.055114$ & $\emph{0.004435}$ \\
\bottomrule
\end{tabular}
\caption{\capcap{Pearson correlations for visual distance.} This table shows the Pearson correlation coefficients between visual distance and either average sale price or number of sales for each of the $9$ case study collections. The final column shows the relationship between visual distance with the last 10 percentiles censored and sale price. All values are statistically significant ($p<0.05$) unless italicized.}
\end{table}

\begin{table}[H]
\begin{tabular}{lrrr}
\toprule
Collection & Sale Price Corr. & \# of Sales Corr. & Censored Sale Price Corr. \\
\midrule
azuki & $-0.170883$ & $0.163149$ & $-0.114551$ \\
beanzofficial & $-0.165244$ & $0.168878$ & $-0.082399$ \\
boredapeyachtclub & $-0.063194$ & $0.111523$ & $-0.031122$ \\
clonex & $-0.183957$ & $0.176747$ & $-0.136650$ \\
cool-cats-nft & $-0.148992$ & $0.281166$ & $-0.097932$ \\
doodles-official & $-0.173598$ & $0.177977$ & $-0.045757$ \\
mutant-ape-yacht-club & $-0.213170$ & $0.209414$ & $-0.141859$ \\
proof-moonbirds & $-0.204417$ & $0.266713$ & $-0.077283$ \\
pudgypenguins & $-0.073887$ & $0.171108$ & $-0.020776$ \\
\bottomrule
\end{tabular}
\caption{\capcap{Pearson correlations for rarity ranks.} This table shows the Pearson correlation coefficients between rarity ranks and either average sale price or number of sales for each of the $9$ case study collections. The final column shows the relationship between rarity ranks with the last 10 percentiles censored and sale price. }
\end{table}

\clearpage

\section{Supplementary Figures}

\label{app:case_study_figs}
\begin{figure*}[h]
    \centering
    \includegraphics[width=\linewidth]{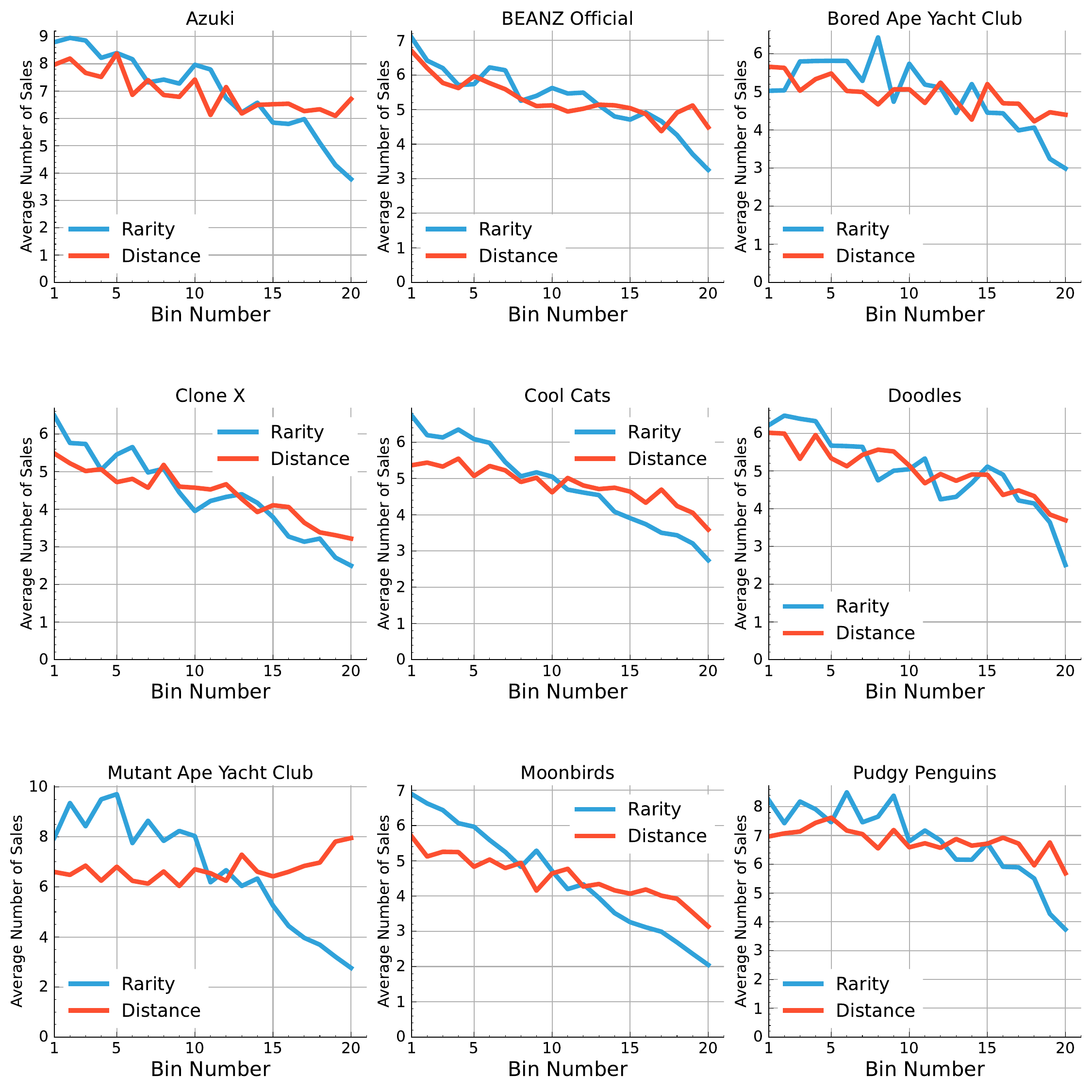}
    \caption{\capcap{Relationships with number of sales.} These plots shows the relationship between rarity rank (visual distance) binned into $20$ quantiles and number of sales of the NFT. In the case of rarity ranks bins are sorted from highest rarity rank (least rare) to lowest rarity rank (most rare).}
    \Description[Relationships with number of sales.]{These plots shows the relationship between rarity rank (visual distance) binned into $20$ quantiles and number of sales of the NFT. In the case of rarity ranks bins are sorted from highest rarity rank (least rare) to lowest rarity rank (most rare).}
    \label{fig:all_slugs_number_sales}
\end{figure*}

\begin{figure*}[ht]
    \centering
    \includegraphics[width=\linewidth]{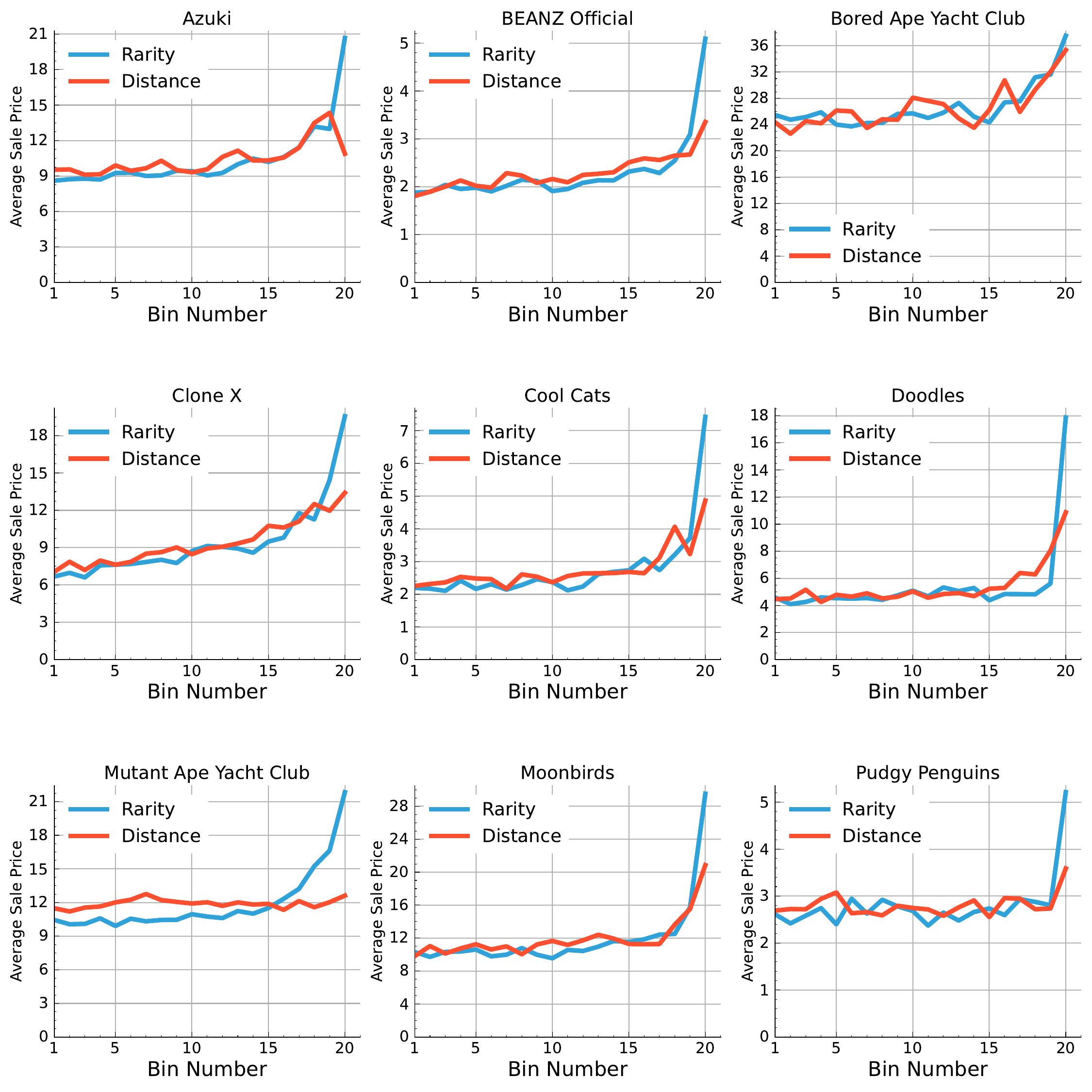}
    \caption{\capcap{Relationships with number of sales.} These plots shows the relationship between rarity rank (visual distance) binned into $20$ quantiles and average sale price of the NFT. In the case of rarity ranks bins are sorted from highest rarity rank (least rare) to lowest rarity rank (most rare).}
    \Description[Relationships with number of sales.]{These plots shows the relationship between rarity rank (visual distance) binned into $20$ quantiles and average sale price of the NFT. In the case of rarity ranks bins are sorted from highest rarity rank (least rare) to lowest rarity rank (most rare).}
    \label{fig:all_slugs_sale_price}
\end{figure*}
\end{document}